\documentclass[useAMS,usenatbib]{mn2e}
\usepackage{times}
\usepackage{url}
\usepackage{graphicx}
\usepackage{amssymb}
\usepackage{amsmath}
\usepackage{natbib}
\usepackage{xspace}
\graphicspath{{./figs/}}

\newcommand\incl{\ensuremath{i}}

\newcommand\rg{\ensuremath{r_\mathrm{g}}\xspace}
\newcommand\risco{\ensuremath{r_\mathrm{ISCO}}\xspace}

\newcommand\rin{\ensuremath{R_\mathrm{in}}\xspace}
\newcommand\rout{\ensuremath{R_\mathrm{out}}\xspace}

\newcommand\fret{\ensuremath{f_\mathrm{ret}}\xspace}
\newcommand\fbh{\ensuremath{f_\mathrm{bh}}\xspace}
\newcommand\finf{\ensuremath{f_\mathrm{inf}}\xspace}

\newcommand\Einc{\ensuremath{E_\mathrm{o}}\xspace}
\newcommand\rinc{\ensuremath{r_\mathrm{o}}\xspace}

\newcommand\Tf{\ensuremath{T_f}\xspace}

\newcommand\Ee{\ensuremath{E_\mathrm{e}}\xspace}

\newcommand\Ne{\ensuremath{N_\mathrm{e}}\xspace}
\newcommand\re{\ensuremath{r_\mathrm{e}}\xspace}

\newcommand\gstar{\ensuremath{g^*}\xspace}

\newcommand\Firrad{\ensuremath{F_\mathrm{irrad}}\xspace}

\newcommand\Femit{\ensuremath{F_\mathrm{e}}\xspace}
\newcommand\Fret{\ensuremath{F_\mathrm{o}}\xspace}
\newcommand\Fsource{\ensuremath{F_\mathrm{p}}\xspace}

\newcommand\xillver{\textsc{xillver}\xspace}
\newcommand\relxill{\textsc{relxill}\xspace}

\newcommand{\dif}[0]{\ensuremath{d}\xspace}
\renewcommand{\o}{_\mathrm{o}}
\newcommand{\e}{_\mathrm{e}}

\title[Returning Radiation]{The Effect of Returning Radiation on Relativistic Reflection}

\author[T. Dauser et al.]{T.\ Dauser$^{1}$\thanks{E-mail:
    thomas.dauser@fau.de}, J.\ A.\ Garc\'ia$^{1,2}$, A.\ Joyce$^{1}$, S.\ Licklederer$^{1}$,
    R.\ M.\ T.\ Connors$^{2}$, A.\ Ingram$^{3,4}$,\newauthor C.\ S.\ Reynolds$^{5}$, J. Wilms$^{1}$ \\ $^{1}$ Dr.\ Karl Remeis-Observatory and
  Erlangen Centre for Astroparticle Physics, Sternwartstr.~7, 96049
  Bamberg, Germany \\ $^{2}$ Cahill Center for Astronomy and Astrophysics, California Institute of Technology, Pasadena, CA 91125, USA  \\ $^{3}$ Department of Physics, Astrophysics, University of Oxford, Denys Wilkinson Building, Keble Road, Oxford OX1 3RH, UK \\
$^{4}$ School of Mathematics, Statistics and Physics, Newcastle University, Herschel Building, Newcastle upon Tyne, NE1 7RU, UK \\ $^{5}$ Institute of Astronomy, University of Cambridge, Madingley Road, Cambridge CB3 0HA, UK } 

\begin{document}

\pagerange{\pageref{firstpage}--\pageref{lastpage}} \pubyear{2021}
\maketitle
\label{firstpage}

\begin{abstract}
We study the effect of returning radiation on the shape of the X-ray reflection spectrum in the case of thin accretion disks. We show that the returning radiation mainly influences the observed reflection spectrum for a large black hole spin ($a>0.9$) and a compact primary source of radiation close to the black hole at height $h<5\rg$, and that it dominates the reflected flux for extreme values of spin and compactness. The main effect of the returning radiation is to increase the irradiating flux onto the outer parts of the accretion disk, leading to stronger reflection and a flatter overall emissivity profile. By analyzing simulated observations we show that neglecting returning radiation in existing studies of reflection dominated sources has likely resulted in overestimating the height of the corona above the black hole. An updated version of the publicly available \relxill suite of relativistic reflection models which includes returning radiation is also presented.
\end{abstract}


\section{Introduction}
\label{sec:introduction}

The reflection of X-rays from the innermost regions of the accretion
disk around black holes, also known as \emph{relativistic reflection}, has been extensively studied in recent decades since it was first postulated \citep{Lightman1988a,Fabian1989} and detected by \citet{tanaka1995a} in the X-ray spectrum of the Active Galactic Nucleus (AGN) MCG$-$6-30-15. Due to the strong gravitational energy shift in the vicinity of the black hole, all reflection features are strongly broadened \citep[see, e.g.,][]{Dauser2010a}, with the strongest being the iron K$\alpha$ emission line at 6.4\,keV. Since the shape of the reflection spectrum is affected by the properties of space-time in the vicinity of the black hole, accurately modeling this is one of the major pathways towards measuring the spin of the black hole \citep[see][for a recent review]{reynolds2020}.

While early studies of relativistic reflection had to concentrate on the broadened iron K$\alpha$ line as the most prominent reflection feature, the advent of high signal-to-noise observations and the broader band-pass of modern instruments such as \textsl{NuSTAR} now permits using the shape of the full reflection spectrum in the measurements \citep[see, e.g.,][]{Dauser2012a,Risaliti2013a,jiang2018}. A large variety of powerful models exist to predict the full and detailed relativistic reflection spectrum, such as \relxill \citep{Garcia2014a,Dauser2014a}, \textsc{reflkerr} \citep{niedzwiecki2019}, the \textsc{KY} package \citep{Dovciak2004a,dovciak2022}, and \textsc{reltrans} \citep{ingram2019,mastroserio2021}. The basic assumption of these  models is that there is a primary source of X-rays, often called the \emph{corona}, which irradiates the accretion disk. This radiation is then reprocessed in the accretion disk \citep[e.g., modeled with \xillver,][]{Garcia2013a}. Relativistic
effects are then imprinted on the reflection spectrum as it is emitted and then propagated to the observer.

The radial dependency of the reflected flux is described by the \emph{emissivity}, which is defined as the emitted bolometric flux. Assuming energy conservation when the radiation is reprocessed, the flux of the reflection component is equal to the irradiating flux. In the simplest approach to reflection modeling, the emissivity is parameterized in the models through an empirical radius-dependent emissivity law, typically a power law.  Using a geometrical model of the corona, such as, e.g., a point source on the rotational axis of the black hole, the irradiation of the disk can be directly determined \citep{Martocchia1996a,Wilkins2012a,Dauser2013a}. The latter \emph{lamp post} configuration naturally produces strongly focused irradiation of the inner accretion disk for a source at low height above the black hole, as is found in many observations \citep{Dauser2013a}. 

One important issue that has so far been almost completely neglected in the modeling of relativistic reflection is the effect of reflected radiation returning to the disk, which will inevitably influence the shape of the reflection features. Here, returning radiation is defined as radiation that is emitted by the accretion disk and then, due to relativistic light bending effects, irradiates other parts of the accretion disk. 

Returning radiation was first studied by \citet{cunningham1976}, who predicted the basic effect of thermal returning radiation contributing to flux emitted by the disk at energies above 10\,keV. More than 20\,years after this initial work, \citet{dabrowski1997} calculated the  effects of returning radiation on relativistic line broadening. Due to their assumptions of a corona co-rotating with the disk and an emissivity of zero at the ISCO, \citeauthor{dabrowski1997} found that the effect of returning radiation was relatively minor. A more general study on returning radiation was presented by \citet{Agol2000a}, who focused on the effect of the returning radiation on the disk structure described initially by a relativistic Novikov-Thorne profile \citep{Novikov1973a,page1974}. \citeauthor{Agol2000a} found that the major effect of the returning radiation is a substantial increase in the flux emitted by the inner regions of the disk. Following this approach, \citet{reynolds2004} presented a model which uses a Novikov-Thorne disk to predict the flux emitted from a corona sandwiching the disk, including additional flux from returning radiation, and applied it to observational data of MCG$-$6-30-15. The effect of returning radiation on the thermal equilibrium of the black body radiation emitted from the disk has been studied in detail and included in, e.g., the \textsc{kerrbb} model \citep{Li2005a}. In a detailed study on how different coronal geometries affect broad relativistic lines, \citet{niedzwiecki2008} also included returning radiation and found that it has the largest effect on the line shape in the lamp post geometry, if the black hole is rapidly rotating and the corona is compact. 

Studies on the effect of returning radiation on the full relativistic reflection spectrum, and not only a single broadened line, are sparse. This is due to the complexity of the implication that returning radiation changes the irradiating spectrum causing the reflection throughout the disk. Basic Monte Carlo studies by \citet{suebsuwong2006} and \citet{riaz2021} use a simplified treatment to calculate the reflection and are limited to neutral accretion disks, which are usually not seen in observations of the inner accretion disk. While the effect of the returning radiation is calculated for one reflection spectrum by \citet{niedzwiecki2016}, it is not included in the \textsc{reflkerr} model \citep{niedzwiecki2019}. In a different Monte Carlo case study, \citet{wilkins2020} calculate the effect of returning radiation on the reflection spectrum and the reverberation measurements using dedicated \textsc{xillver} \citep{Garcia2013a} reflection tables. Due to the complexity of this approach, \citeauthor{wilkins2020}  presented results for only one parameter combination and only using an averaged energy shift. The latter assumption leads to biased results due to the importance of the strong energy shifts inherent in returning radiation \citep{dabrowski1997}. Lastly, \citet{connors2020} presented the first evidence for the existence of returning radiation from the thermal disk radiation that produces the relativistic reflection in the Black Hole X-ray Binary XTE~J1550$-$564.

Despite the improvement in measurements and models for relativistic reflection in the last two decades however, no general effort has been made to predict the effect of the returning radiation on the reflected spectra where relativistic reflection is typically observed. In this paper we present the first detailed study of returning radiation that produces relativistic reflection along with a model implementation published in the publicly available \relxill-code. We ignore the disk-intrinsic emission and focus purely on reflection dominated sources, where radiation from the corona induces the primary reflection that will then be returning to the accretion disk. The basic ray-tracing setup for calculating the returning radiation is described in Sect.~\ref{sec:simulation}, which is then applied in Sect.~\ref{sec:observed-spectrum} to predict the observed spectrum. The results are presented in Sect.~\ref{sec:results}. The overall implications of our results are discussed in Sect.~\ref{sec:discussion} and then summarized and concluded in Sect.~\ref{sec:summary-conclusions}.
 
\section{Calculating the Returning Radiation} \label{sec:simulation}

In order to calculate the effect of returning radiation, we perform ray-tracing simulations of single photon trajectories in the Kerr metric for a grid of radii \re\ covering the whole accretion disk. For a given dimensionless spin value, $a$, a fraction of these trajectories will be bent back onto the disk at $\rinc$. Assuming isotropic emission in the emitter frame, the total returning radiation at $\rinc$ is then obtained by summing the contributions from all disk radii, taking special and general relativistic effects into account. In the following, we describe this computation in detail.

\subsection{Ray Tracing} \label{sec:ray-tracing}

For a given radius of emission $\re$, we simulate a large number of photon trajectories with their initial directions emitted isotropically in the fluid frame from the flat surface of the accretion disk \citep{Agol2000a}. This means that the photon flux from a surface element $\dif A$ is proportional to 
\begin{equation} \label{eq:fluid-frame}
    I\cos(\theta)\dif\Omega=0.5I\sin(2\theta)\dif\theta \dif\phi
\end{equation}
 \citep{chandrasekhar1960}. We use the YNOGK code \citep[][hereafter YW13]{yang2013} to calculate the null geodesics, which describe the photon trajectories (see Appendix~\ref{sec:appendix-ray-tracing} for the basic ray-tracing equations). The accretion disk, which is assumed to be razor-thin, is described by circular particle orbits in the equatorial plane in the Kerr metric \citep[and Appendix~\ref{sec:appendix-ray-tracing}]{Bardeen1972}.  The input to the YNOGK code is the 3-velocity $V^{(\phi)}(\re)$ of the emitter\footnote{defined with respect to the locally non-rotating frame (LNRF)}, i.e. in our case the accretion disk. The initial photon momentum is calculated from the direction of emission in $\theta$ and $\phi$ in the fluid frame, as defined in YW13.

Due to the axis-symmetric nature of the metric and the accretion disk (i.e., symmetry in $\phi$ direction), the problem is fully described by photons emitted at $\re$ at the coordinate $\phi=0$. For each of these photon trajectories we calculate the incident radius \rinc~and the azimuthal $\phi$ coordinate at the point of incidence. The $\phi$ dependence is important as, due to the circular velocity of the accretion disk, the photons will experience an energy shift which depends on both the incident radius and on $\phi$. In other words, seen from the radius of emission, a part of the ring where the photons are incident (\rinc) will always move towards the observer, while another part will recede. 
  
\begin{figure*}
  \centering
  \raggedleft
  \includegraphics[width=0.97\textwidth]{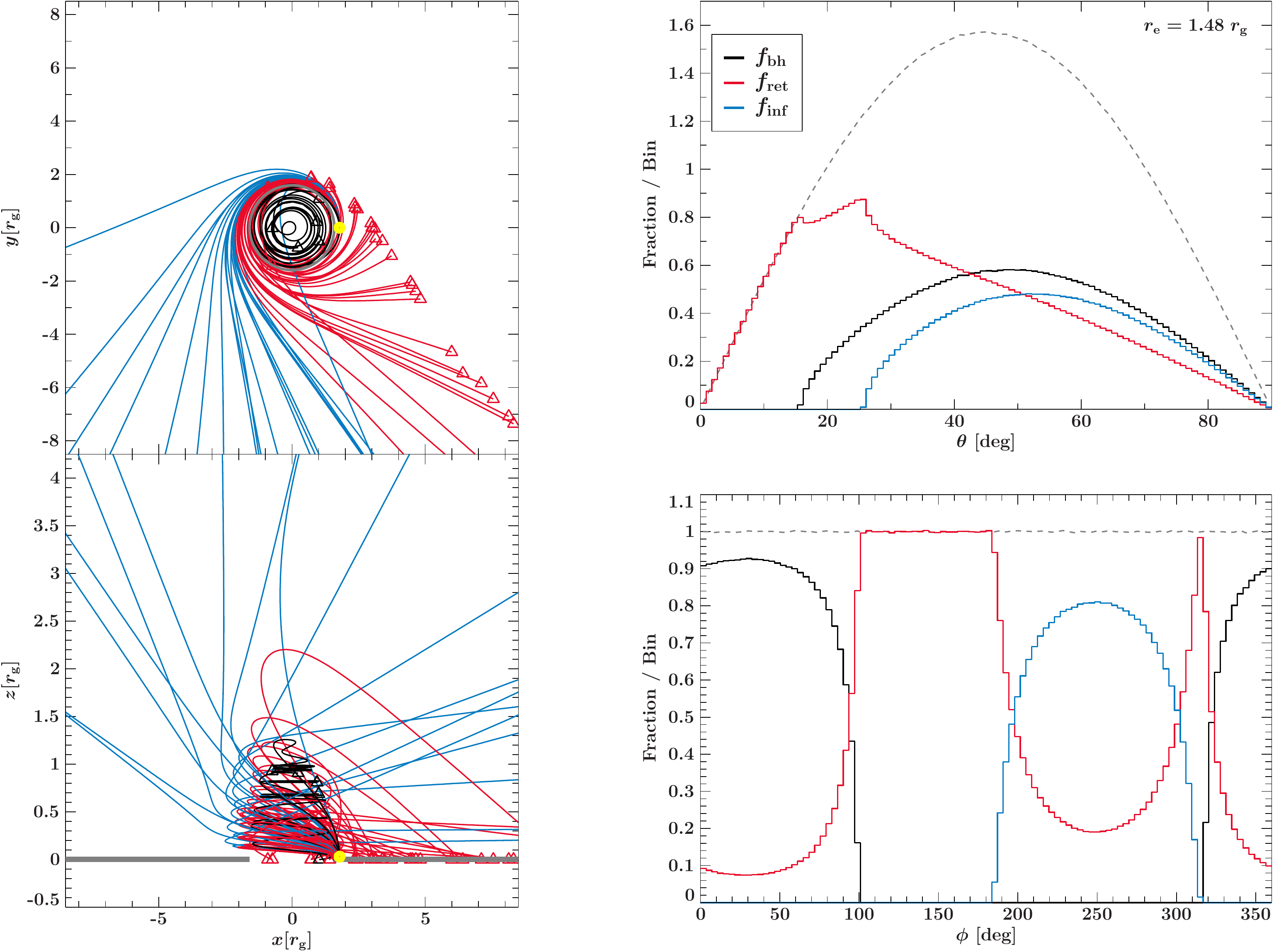}
  \caption{(Left) Photon trajectories for spin $a=0.998$ emitted at $\re=1.48\rg$ (yellow circle). Triangles indicate the location where the photons hit the disk (red) or the black hole (black). Photons reaching infinity are drawn in blue. (Right) Distribution of photons depending on $\theta$ and $\phi$ (in the frame of the disk), color coded by where they will end up. $\theta$ is defined such that $\theta=0^\circ$ is perpendicular to the accretion disk plane. As seen from the point of emission, the rotation of the disk is in direction of $\phi=0^\circ$ and therefore the black hole is located in direction of $\phi=90^\circ$. Note that we use an isotropic source in the fluid frame and therefore the fraction of photons depends on $\theta$, as indicated by the gray line, which shows the sum of all photons for a given value of $\theta$ and $\phi$.}
  \label{fig:dist-single-rays}
\end{figure*}

Figure~\ref{fig:dist-single-rays} shows an example of photon trajectories emitted isotropically from one single radius very close to the black hole. It is immediately evident that extreme light-bending takes place in this strong gravity regime, with photons circling around the black hole in its direction of rotation. The right panel of Fig.~\ref{fig:dist-single-rays} shows the fraction of photons as a function of the emission angles $\theta$ and $\phi$. The angle $\theta$ is defined with respect to the normal of the disk surface. As can be seen, for larger angles towards the disk (smaller values of $\theta$), more photons are returning to the disk. Photons that are emitted more parallel to the disk surface are more likely to fall into the black hole or to escape to infinity. The azimuthal emission angle, $\phi$, strongly determines where the photon will end up. It is defined in the plane of the accretion disk, such that for $\phi=0^\circ$ the photon is emitted in the direction of movement of the disk. For such a close vicinity to the black hole, only photons emitted directly away from the black hole ($\phi=270^\circ$) can escape. The majority of photons falling into the black hole are emitted in the direction of motion of the disk ($\phi=0^\circ$). Most photons returning to the disk are emitted in the opposite direction of the disk rotation. Therefore, the fate of emitted photons is decided by the emission angle $\phi$; whether they are lost to the black hole ($\phi < 90^\circ$) or return to the disk ($\phi > 90^\circ$). 

Using the simulated trajectories, we can determine the fraction of isotropically emitted photons that will return to the disk, \fret, reach infinity, \finf, or fall into the black hole, \fbh. Following the reasonable definition of \citet{Agol2000a}, we count a photon as returning to the disk if it crosses the equatorial plane at a radius that is larger than the innermost stable circular orbit (ISCO)\footnote{For computational purposes we choose an outer radius of $10^5\rg$, although in reality a small fraction of photons will hit the disk beyond this value.}. If a photon does not directly hit the event horizon, but crosses the plane at $r<\risco$, we count it as being captured by the black hole, as we assume \citep[analogous to][]{Agol2000a} that it will be advected or scattered inwards by the in-flowing matter.

\begin{figure}
  \centering
  \includegraphics[width=\columnwidth]{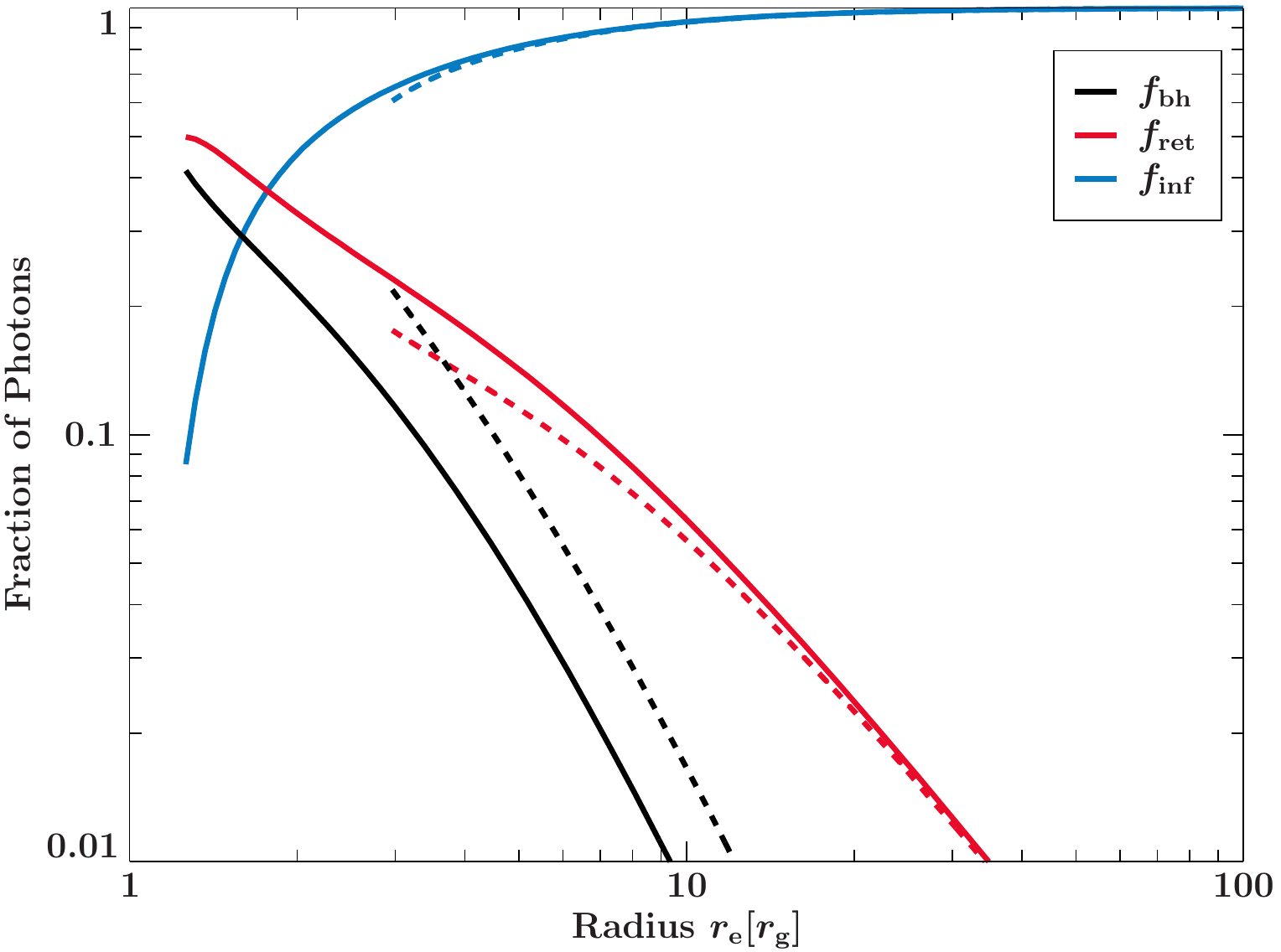}
  \caption{Fraction of photons falling into the black hole, returning to
    the disk, or reaching infinity as a function of radius of emission,
    $r_\mathrm{e}$ for the case of a maximally spinning black hole
    ($a=0.998$, solid lines) and for $a=0.8$ (dashed).
    Photons are emitted isotropically (in the frame of rest of the
    disk) from the flat surface of the  accretion disk.}
  \label{fig:photon_fate}
\end{figure}

Figure~\ref{fig:photon_fate} shows the fraction of photons either returning to the disk, falling into the black hole, or reaching infinity. For a rapidly rotating black hole with $a=0.998$, at the ISCO the fraction of photons returning to the disk (${\sim} 50\%$) is similar to that being captured by the black hole (${\sim} 40\%$), while only a small fraction escapes to infinity. For increasing radius, the fraction of photons returning to the disk drops to only ${\sim} 10\%$ at $10\rg$, and it is even less beyond this radius. For a more slowly rotating black hole the ISCO gets larger, and consequently the fraction of photons intercepted by the disk decreases. In detail, however, the fraction of photons reaching infinity for a given radius of emission is largely independent of spin, which mainly determines the ratio of the fraction of photons returning to the disk to that captured by the black hole. A smaller spin, and therefore a larger ISCO, increases the fraction of photons being captured by the black hole (see Fig.~\ref{fig:photon_fate}). In this case the majority of the photons are already reaching infinity, however, such that the observable effect of this spin dependency is small.

\subsection{Returning Flux and Transfer Function}
\label{sec:returning-flux-disk}

In order to determine the radiation returning to the accretion disk, we perform the ray-tracing calculations outlined above for a grid of radii \re\ covering the whole disk. Following \citet{Cunningham1975}, we can write the returning photon flux $[\mathrm{photons}/\mathrm{s}/\mathrm{cm}^{2}/\mathrm{keV}]$ as measured at $\rinc$ as the integral
\begin{equation}\label{eq:Fret}
  \Fret(\Einc,\rinc) =  \int_{\rin}^{\rout} \int_0^1
  \frac{\Tf(\rinc,\re,g^*)}{\re} \Femit(\Einc/g,\re) \; \dif g^*\, \dif\re
\end{equation}
where $\Tf$ is the \emph{flux transfer function}. We parameterize the $\phi$ coordinate directly in terms of the energy shift $g$, as is commonly done in such ray tracing calculations \citep[see, e.g.,][]{Cunningham1973}. This energy shift $g$ of photons emitted at \re\ and observed at \rinc\ is defined as
\begin{equation}\label{eq:ener-shift}
  g = \frac{\Einc}{\Ee} = \frac{p_\mu u(\rinc)^\mu}{p_\mu u(\re)^\mu} \quad,
\end{equation}
where $p_\mu$ is the four-momentum of the photon and $u^\mu$ is the \mbox{four-velocity} of the particles in the accretion disk (see Appendix~\ref{sec:appendix-ray-tracing}). For ease of computation, we use the dimensionless energy shift, $g^*$, which is defined as
\begin{equation}
  g^* = \frac{g - g_\mathrm{min}}{ g_\mathrm{max} - g_\mathrm{min}} \quad.
\end{equation}
The quantity $\Femit(E\e,\re)$ in Eq.~\ref{eq:Fret} describes the specific emitted photon flux in the frame of $\re$, which depends on the incident flux of the primary source, \Fsource. In general, as the radiation emitted at $\re$ is produced by reprocessing in the accretion disk atmosphere, detailed radiative transfer calculations are required to correctly calculate $\Femit(E\e,\re)$ \citep[e.g., by using \xillver;][]{Garcia2013a}. 

The fundamental quantity encoding the light-bending of the returning radiation is the dimensionless transfer function \Tf, given by
\begin{equation}
 \Tf(r\e, r\o, g^*) = \frac{\mu\e }{\pi g} \frac{r\e^2}{r\o}~
        \left|\frac{\partial \Omega\e(r\o, g^*)}{ \partial(r\o, g^*)} \right| \;,
\end{equation}
where $\mu\e=\cos\theta\e$ and where $\dif\Omega\e = [\partial\Omega\e/\partial(r\o, g^*)]dr\o dg^*$ is the solid angle under which photons emitted at $r\e$ are incident on the annulus with radius $r\o$ and width $\dif r\o$ and with an energy shift $g^*$ within $\dif g^*$ (see Appendix~\ref{sec:appendix-tf} for a detailed derivation, and Appendix~\ref{sec:append-invariance} for a discussion on an inconsistency in previous approaches that happens to still give the correct answer for primary reflection in the lamppost geometry). 

To calculate \Tf, we simulate an equal number of photons $N_\mathrm{tot}(\re)$ for each radial bin \re. As described in Sect.~\ref{sec:ray-tracing}, those photons are distributed isotropically in the fluid frame (see Eq.~\ref{eq:fluid-frame}). We then count the number of photons, $\Delta N(\rinc,\re,g^*)$, that impact the disk at an annulus at radius \rinc with a width $\Delta \rinc$ and in the energy shift bin $\Delta g^*$. Using the fact that those photons are emitted under the solid angle $d\Omega_e$, we can connect it to the fraction of returning photons by
\begin{equation} \label{eq:DeltaN}
  \frac{\Delta N}{N_\mathrm{tot}} = \mu\e \frac{\dif{\Omega_\mathrm{e}}}{\pi} \quad. 
\end{equation}
Note that the $\mu\e$ factor takes into account that in our ray-tracing calculation the photons are already distributed isotropically in the fluid frame (see Sect.~\ref{sec:ray-tracing}). The discretized transfer function is then given by
\begin{equation}\label{eq:tf}
  \Tf(\rinc,\re,g) = 
  \frac{r_\mathrm{e}^2}{g~r_\mathrm{o}} \cdot
  \frac{\Delta N(\rinc,\re,g^*)}{N_\mathrm{tot}(\re) }  \cdot
  \frac{ 1 }{\Delta r_\mathrm{o} \Delta g^*} \quad.
\end{equation}
We note that the complexity of calculating the transfer function is  hidden in the calculation of $\Delta N(\rinc,\re,g^*)$, which is done by simulating photon trajectories using the YNOGK code (YW13) as described in Sect.~\ref{sec:ray-tracing}. This means that instead of using a flux-based approach \citep[see, e.g.,][]{cunningham1976}, we use isotropically distributed photon trajectories to calculate the transfer function. Also note that while our calculations inherently employ this isotropic emission characteristic, a different distribution can be easily taken into account by adding an appropriate $\mu\e$-dependent factor in Eq.~\ref{eq:DeltaN}.

\begin{figure}
  \centering
\includegraphics[width=\columnwidth]{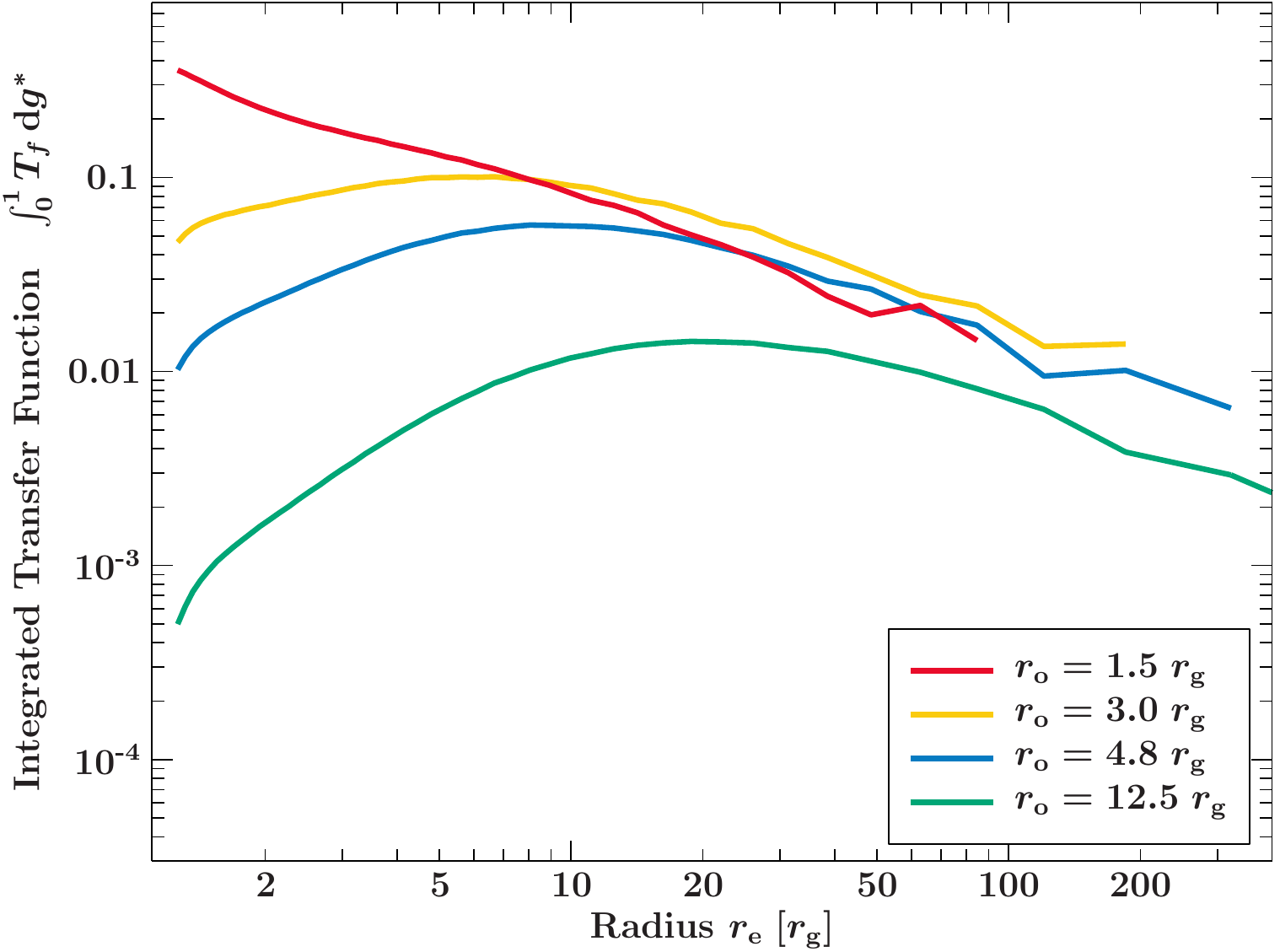}
  \caption{Integrated returning radiation transfer function for a few selected values of \rinc\ as a function of \re.  }
  \label{fig:transfer_function}
\end{figure}

Figure~\ref{fig:transfer_function} shows the transfer function \Tf for a few selected values of incident radius \rinc. We show it integrated over the energy shift in order to visualize its dependence on $\re$. The figure shows that it is a smooth and slowly varying function of the emission radius \citep[see also,][]{cunningham1976}. The largest differences are evident for the smallest emission radii, where the value of the transfer function is two magnitudes larger for a smaller incident radius \rinc compared to a larger one. 

\begin{figure}
  \centering
\includegraphics[width=\columnwidth]{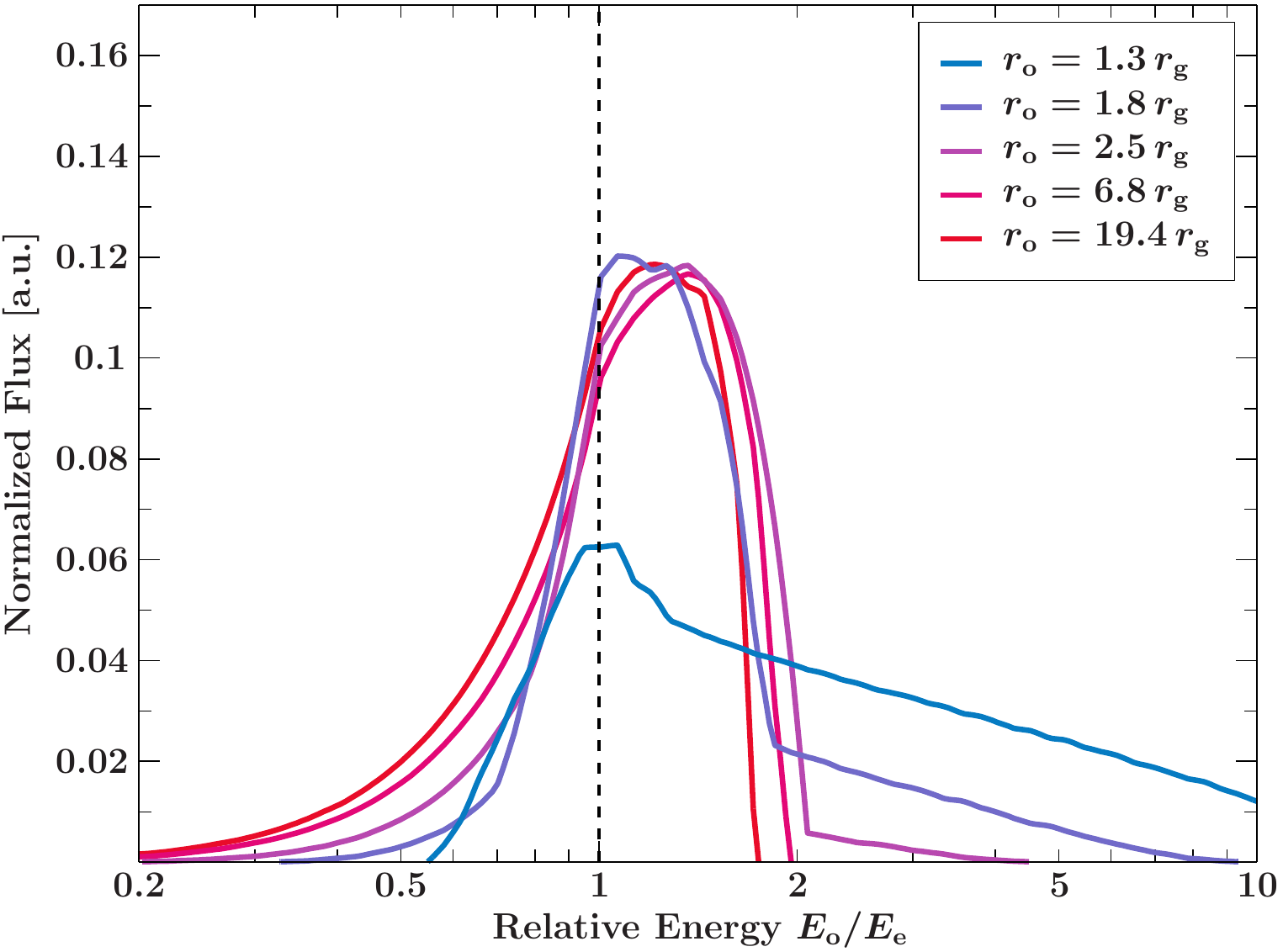}
  \caption{Normalized spectrum of mono-energetic radiation emitted at $\Ee$ over the whole disk, plotted as function of the relative energy $\Einc/\Ee$ (identical to the  energy shift $g$) as observed at certain radii \rinc on the disk. We use the standard emissivity of $\varepsilon \propto r^{-3}$ and set the spin to $a=0.998$. }
  \label{fig:gshift_line}
\end{figure}

The combined effect of the energy shift and the transfer function can be visualized by calculating the returning spectrum for a mono-energetic emission line emitted over the whole disk at an energy $\Ee$. Figure~\ref{fig:gshift_line} shows the returning line profile as observed at selected radii \rinc on the disk for the standard $r^{-3}$ emissivity expected from a \citet{Shakura1973a} disk. The spectra have been re-normalized to the same observed flux to visualize the difference in spectral shape. Overall, it can be seen that the returning radiation experiences a strong energy shift, in the range of $g=0.7$--2. Despite the overall similarity in line shape, a closer look reveals differences of the returning spectrum depending on the point of observation. At the smallest radii, a fraction of photons experiences a large blueshift (up to a factor of 10). These are photons emitted at the outer regions of the disk and returning to very close to the black hole, thus experiencing a large gravitational blue shift from falling further into the gravitational potential well. For an increasing radius, the amount of redshifted photons increases, while at the same time the majority are still blueshifted by almost a factor of two.

\subsection{Irradiating Flux and Emissivity Profile}
\label{subsec:emissivity-profile}

The emissivity profile plays a fundamental role in modeling relativistic reflection. It characterizes the radial dependence of the reflected flux from the accretion disk. As relativistic effects depend strongly on the radius, the emissivity has a major influence on relativistic reflection. Additionally, it is either a direct input to relativistic reflection models or it can be easily predicted for specific primary source geometries such as the lamp post \citep[see, e.g., ][]{Dauser2013a}. In the following we will calculate how returning radiation changes the overall emissivity profile. 

In the context of relativistic reflection, the emissivity profile is defined as the radial dependent flux reflected from the accretion disk in the energy band relevant for X-ray reflection. It is usually denoted by $\varepsilon(r)$. Furthermore, it is typically assumed that the flux is conserved when the primary radiation is reflected at the disk. Therefore the emissivity profile also describes the radial dependency of the incident flux. 

Including returning radiation, the specific flux irradiating the accretion disk at radius $\rinc$ is the sum of the direct incident flux from the primary source, $\Fsource(\Einc, \rinc)$, and the flux of the radiation returning from all parts of the accretion disk, $\Fret(\rinc)$, that impinges on the disk at $\rinc$, leading to
\begin{equation}\label{eq:Femit}
  F(\Einc, \rinc) = \Fsource(\Einc, \rinc) + \Fret(\Einc, \rinc) \quad. 
\end{equation}
Note that we neglect any contribution to the returning radiation from thermal emission emitted from the disk, as well as any change in the disk temperature caused by the irradiating radiation. 

The emissivity profile at a given radius $r\o$ can now be calculated by integrating the irradiating specific flux. The integration is performed over a fixed energy band that we assume is relevant for X-ray reflection. Including returning radiation, the emissivity is therefore given as
\begin{align} \label{eq:emis-general}
\varepsilon(r\o) &= \int_{E_\mathrm{lo}}^{E_\mathrm{hi}} F(E\o,r\o) dE\o  \\ \label{eq:emis-total}
   &= \int_{E_\mathrm{lo}}^{E_\mathrm{hi}} \left[ \Fsource(\Einc, \rinc) + \Fret(\Einc, \rinc) \right] dE\o 
  \\
  &=: \varepsilon_\mathrm{p}(r\o) + \varepsilon\o(\varepsilon_\mathrm{p}, r\o)
  \quad, 
\end{align}
where $\varepsilon_\mathrm{p}(r\o)$ is the emissivity of primary reflection from direct irradiation of the primary source. The contribution of returning radiation, $\varepsilon\o(\varepsilon_\mathrm{p}, r\o)$, depends also on the primary emissivity, which determines the amount of returning flux.

The reflected spectrum $\Femit(\Ee, \re)$ that will be returning to the disk can then be separated in an energy dependent spectral shape $N\e(\Ee)$ and the radial dependent emissivity $\varepsilon_\mathrm{p}(r\e)$ 
\begin{equation}\label{eq:emis-spectral-shape}
 F_\mathrm{e}(E\e,r\e) = N_\mathrm{e}(E\e) \varepsilon_\mathrm{p}(r\e) \quad.
\end{equation}
Following the definition of the emissivity in Eq.~\ref{eq:emis-general}, the spectral shape $N\e$ is defined such that it is normalized according to
\begin{equation}\label{eq:emis-norm}
 \int_{E_\mathrm{lo}}^{E_\mathrm{hi}} N\e(E\e) dE\e = 1 \quad.
\end{equation}
We note that this normalization also holds in the frame of $E\o$ for the observed spectrum $N\o(E\o)$ for the same energy band, as any change in flux is by definition absorbed in the emissivity as defined in Eq.~\ref{eq:emis-spectral-shape}.

To calculate the emissivity by integrating $\Fret(E\o,r\o)$ (see Eq.~\ref{eq:emis-total}), we need to know the reflected spectrum $\Femit(\Ee, \re)$ that will be returning to the disk. This reflection spectrum, however, depends on many parameters like the density and ionization of the accretion disk, or the incident spectrum \citep[e.g., ][]{Garcia2013a}. 
To study the effect returning radiation has on the emissivity profile, we therefore assume that the disk acts as a ``perfect reflector'', meaning the reflected spectrum is equal to the irradiating spectrum\footnote{The validity of this assumption strongly depends on the ionization of the reflector and the photon index $\Gamma$ of the irradiating source spectrum, but serves to get a basic understanding of the effects of the returning radiation. For an extended discussion see Sect.~\ref{sec:new-relat-refl-model} and Appendix~\ref{sec:flux-corr-fact} explaining possible correction factors to this approximation, which are used to implement returning radiation in the \relxill model.}. 
As the incident spectrum is a power law, the emitted spectrum returning to the disk also follows a power law. 

We can now solve the integral in Eq.~\ref{eq:emis-total}. As the emissivity is calculated in the frame of $r\o$, the integration is performed over $dE\o$ and therefore the spectral shape needs to be transformed from the emitter to the observer frame. As the emitted radiation follows a power law shape, $N_\mathrm{e}(E\e) \propto E^{-\Gamma}$, the normalization of the power law transforms under energy shift as  $ \Ne(\Ee ) = g^\Gamma N\o(\Einc)$ \citep[see also][]{Dauser2013a,ingram2019}. This transformation is consistent with our definition of the normalization in Eq.~\ref{eq:emis-norm} over fixed energy band and leads to 
\begin{equation}
  \Femit(\Ee, \re) = g^\Gamma \Ne(\Einc) \varepsilon_\mathrm{p}(\re) \quad.
\end{equation}
Inserting this in Eq.~\ref{eq:Fret} results in the emissivity profile
\begin{equation}\label{eq:emis-rrad}
  \varepsilon_o(\rinc) =  \int_{R\mathrm{in}}^{R\mathrm{out}} \varepsilon_p(\re) \left[
  \int_0^1 \frac{\Tf(\rinc,\re,g)}{\re}  g^\Gamma \dif g^*\right]  \dif \re \quad.
\end{equation}

\begin{figure*}
  \centering
  \includegraphics[width=\textwidth]{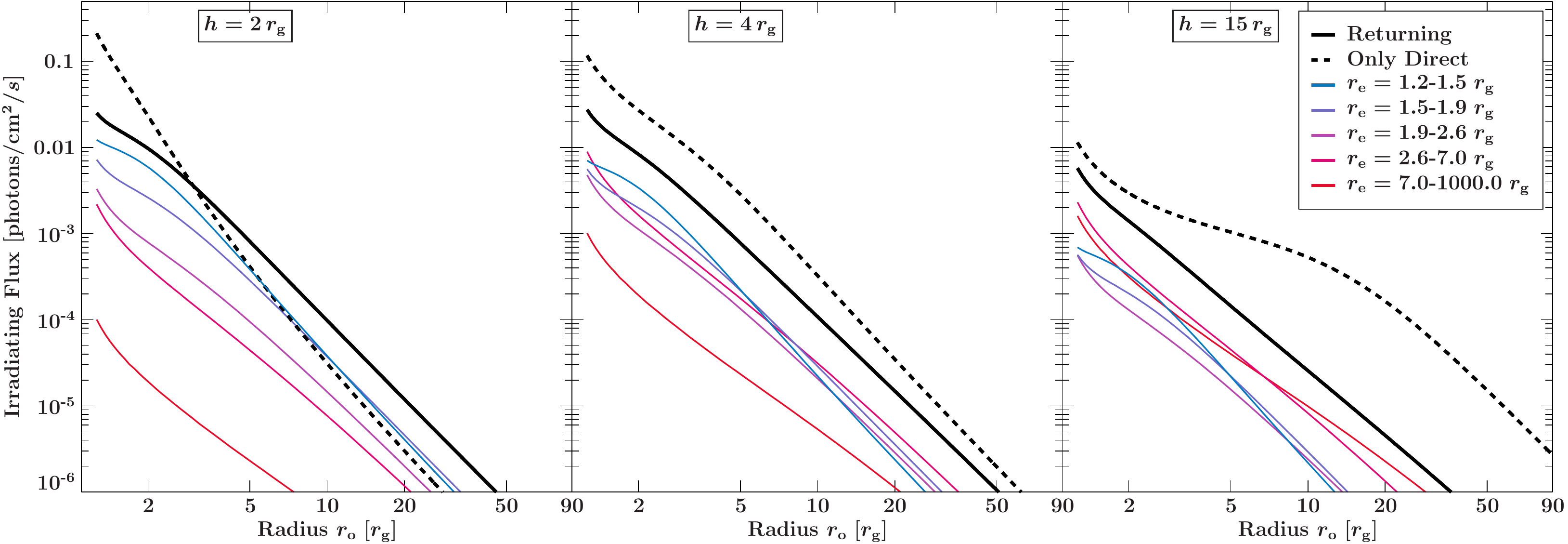}
  \caption{Integrated irradiating flux produced by the returning radiation (solid black line) and the direct incident radiation (dashed line). Color-coded is the returning radiation as emitted from specific regions of the accretion disk, for a spin $a=0.998$, $\Gamma=2$, and for three different  heights. }
  \label{fig:emis_profile_zones}
\end{figure*}
Figure~\ref{fig:emis_profile_zones} shows the irradiating flux created
by the returning radiation from different regions of the disk, assuming the lamp post geometry. In this geometry the source of primary radiation is located at a height, $h$, above the black hole on its rotational axis \citep{Martocchia2000a,Dauser2013a}.  The three different source heights depicted in Fig.~\ref{fig:emis_profile_zones} cover the range of typical irradiation profiles from steep (low height) to flat (large height). In general, the effect of the returning radiation (solid line) is strongest for a small lamp post height. For $h=2\rg$, the returning radiation dominates the emissivity for radii $>3\rg$. The majority of the returning radiation hitting the outer regions of the disk originates from the very inner disk $<2\rg$ (blue curves). While the returning flux does not change much when increasing the source height to $4\rg$, the directly incident flux from the lamp post corona to the outer parts increases. The combination of both effects means that already at $h=4\rg$ the direct irradiation is dominating the irradiation profile with a smaller contribution of the returning radiation (30--50\%, middle panel of Fig.~\ref{fig:emis_profile_zones}). Note that a lamp post source with an intermediate height of $h=4\rg$ produces an emissivity profile that is similar to  the canonical $\varepsilon \propto r^{-3}$ expected from a corona directly coupled to a standard \citet{Shakura1973a} disk\footnote{Note that for such an $\alpha$-disk this profile actually flattens at the inner disk. Assuming zero-torque at the ISCO, the emissivity would follow $[1-\sqrt{(r_\mathrm{ISCO}/r)}]\cdot r^{-3}$.} and therefore can be taken as reference for the effect of the returning radiation in this case.

For a larger source height of $15\rg$ (right panel of Fig.~\ref{fig:emis_profile_zones}), the returning flux is reduced and is only a minor contribution to the overall irradiating flux. Contrary to the small source heights, in this case the radiation is returning from the outer disk ($>3\rg$, red curves for the largest \re dominate the irradiation profile) to the very inner disk ($<3\rg$), and therefore only the very inner radii are affected by returning radiation. Still, while only a very small fraction of the flux is returning from such large radii (see Fig.~\ref{fig:photon_fate}),  the strong blueshift (up to a factor
10, Fig.~\ref{fig:gshift_line}) of the radiation on its way to the
inner disk, and the accompanying flux boost, means that for a source height of $h=15\rg$ the outer disk contributes to 50\% of the flux incident at the inner disk. 

The total observed emissivity,  $\varepsilon_\mathrm{o}$, is composed of the direct radiation plus all contributions from returning radiation to the disk. The latter potentially also includes radiation returning to the disk multiple times. As the returning radiation is mainly influencing the outer parts of the disk (see Fig.~\ref{fig:emis_profile_zones}), however, the effect of higher-order returning radiation will be minor \citep[see also][]{wilkins2020}.

\begin{figure}
  \centering
  \includegraphics[width=\columnwidth]{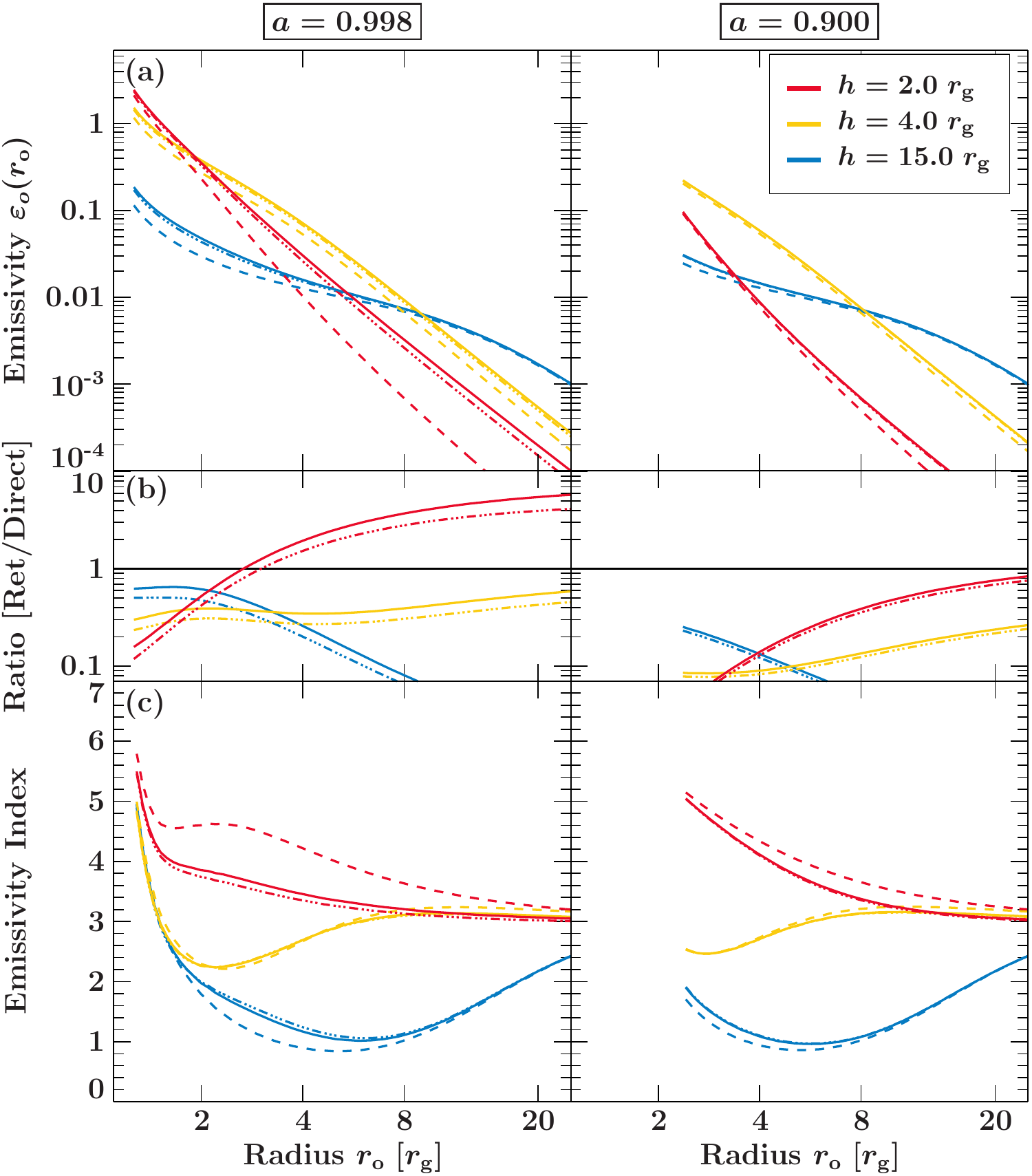}
  \caption{Emissivity profiles including returning radiation for different lamp post source heights. The solid line is including first and second-order returning radiation, while the dashed-dotted line is only including first order returning radiation. For reference, the dashed line shows the emissivity profile without returning radiation. (a) Total emissivity profiles. (b) Ratio between the returned and direct emissivity profile. (c) Emissivity index, defined as $r^{-q}$. For all curves, the spin was set to  $a=0.998$ (left column) and $a=0.9$ (right column) and the power law index to $\Gamma=2$.}
  \label{fig:pl_emis_profile_total}
\end{figure}

Figure~\ref{fig:pl_emis_profile_total}a shows these total emissivity profiles including returning radiation (solid lines). Comparing them to only the direct emissivity (dashed lines) shows the effect of the returning radiation in changing the irradiation of the accretion disk. As already discussed above, for a very low source height of $2\rg$ and maximum spin, $a=0.998$, the emissivity profiles are dominated completely by returning radiation at larger radii. For very small heights, the flux at large radii can be increased by more than a factor of five (Fig.~\ref{fig:pl_emis_profile_total}b) compared to the case where returning radiation is ignored. Even for a height of 4\rg, the returning radiation already plays less of a role, such that the total emissivity more closely resembles the direct emissivity. While returning radiation adds 30-50\% of flux to the direct emissivity, its contribution is comparatively constant over the whole disk (yellow line in Fig.~\ref{fig:pl_emis_profile_total}b) and therefore only leads to an overall increase in emissivity, while the profile itself stays the same. In the case of a lower black hole spin value, $a=0.9$, the returning radiation plays only a minor role in contributing to the direct emissivity profile.

Studies of the properties of reflection often use the \emph{emissivity index} $q$ to characterize the steepness of the emissivity profile, which is often used as a fit parameter when modeling observational data. The emissivity index is derived by approximating the emissivity profile with a power law $\varepsilon \propto r^{-q}$. Figure~\ref{fig:pl_emis_profile_total}c shows the behavior of $q$ when returning radiation is taken into account. It shows that for a very low source heights ($h=2\rg$), returning radiation leads to a strong flattening of the emissivity profile.  For larger source heights, the emissivity index is not affected much by returning radiation. Generally, it can be seen that this flattening leads to more similar emissivity profiles for compact primary sources, which is well described by $r^{-3}$ for radii larger than $5\rg$. For a smaller spin of $a=0.9$, the emissivity profiles are less affected by returning radiation and are therefore even steeper for $h=2\rg$ than those obtained for extreme spin values. 

Lastly, Fig.~\ref{fig:pl_emis_profile_total} also shows the effect of including higher-order returning radiation (solid lines) in comparison with only first-order returning radiation (dashed-dotted lines). As shown in the bottom panel of the figure, for the largest spin it has a net ${\sim}10\%$ increase in emissivity over all radii, and is not at all visible for $a=0.9$. It is only weakly dependent on the radius, and overall the effect is minor compared to the initial returning radiation.

Note we have chosen to use an intermediate and common value of $\Gamma=2$ throughout the paper. While a detailed assessment of the effect of $\Gamma$ is beyond the scope of this publication, the general effect can be understood by looking at the dependence of the lamp post emissivity profile on $\Gamma$, which shows that a larger value of $\Gamma$ will generally increase the irradiation of the inner disk \citep[see, e.g.,][]{Dauser2013a}. Therefore radiation returning from the inner to the outer disk will also be increased. Furthermore, as shown in Eq.~\ref{eq:emis-rrad}, the returning radiation emissivity profile directly depends on $g^{\Gamma}$, which means that a larger value of $\Gamma > 2$ increases the emissivity, as returning radiation is on average blueshifted (see Fig.~\ref{fig:gshift_line}). Using the same arguments as above, a flatter primary spectrum ($\Gamma < 2$) then has the opposite effect on the emissivity profile. 

\section{Observed Reflection Spectrum}
\label{sec:observed-spectrum}

To obtain the shape and flux of the observed (relativistically smeared) reflection spectrum, we first need to calculate the total incident spectrum $\Firrad(E,\rinc)$ on the disk, which is the combination of primary source spectrum, $\Fsource$, and the returning reflected emission $\Fret$ from the whole disk (Eq.~\ref{eq:Femit}). Due to the strongly varying energy shifts between the points of emission and incidence (see Eq.~\ref{eq:ener-shift} and Eq.~\ref{eq:Fret}), the returning spectrum is expected to be relativistically broadened, implying that the combined direct emission and returning spectrum depends on the incident radius \rinc\ in a complex way. Knowing the total incident spectrum, the reflection can then be obtained from radiative transfer calculations \citep[\xillver,][]{Garcia2013a}. In a second step, this total reflected spectrum is convolved with a relativistic kernel in order to take into account the special and general relativistic effects affecting the radiation on its way to the observer. While the latter is a straight-forward convolution, using, for example \relxill\ \citep{Dauser2010a,Garcia2014a}, the main problem lies in the calculation of the reflection of the combined incident spectrum $\Firrad$.

\subsection{Incident Returning Reflection Spectrum}

The spectrum of the reflected radiation returning to the disk as observed at \rinc, $\Fret(E,\rinc)$, is fully defined through Eq.~\ref{eq:Fret}. It is an integral over the primary reflected spectra $\Femit(\Einc, \re)$, weighted with the transfer function and shifted in energy. This means, as is the case for directly observed reflection, the returning spectrum is getting relativistically smeared while returning to other parts of the disk.

\begin{figure*}
  \centering
\includegraphics[width=\textwidth]{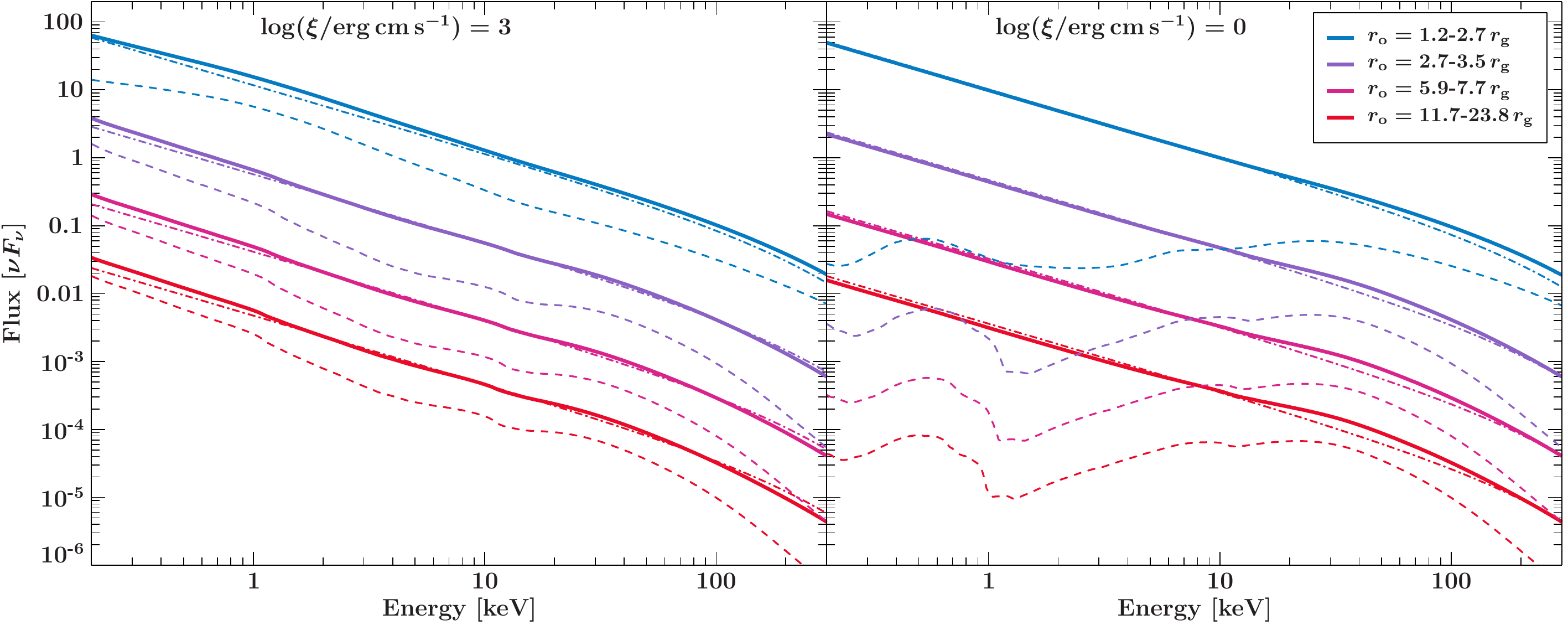}
  \caption{Relativistically blurred returning spectrum as seen by different zones on the disk for a primary source with a photon index $\Gamma=2$, height $h=3\rg$, and a black hole spin of $a=0.998$. (Left) a highly ionized disk with $\log(\xi/\mathrm{erg\,cm\,s}^{-1})=3$ and (right) a neutral disk with $\log(\xi/\mathrm{erg\,cm\,s}^{-1})=0$. Each radial zone is chosen in size such it will roughly contribute an equal amount of flux in the observer frame. The thick solid line shows the combined direct and returning spectrum as seen by this radial zone. It is the sum of the returning spectrum (dashed line) and the direct incident  spectrum (not shown). For comparison, the dashed-dotted line shows the combined direct and returning spectrum using the assumption of approximating the returning reflection as a power law, including a flux correction factor, $C_F$, that ensures local conservation of the flux.}
  \label{fig:relat_reflect_returnrad}
\end{figure*}
Figure~\ref{fig:relat_reflect_returnrad} shows the returning spectrum as incident on certain selected radial zones of the accretion disk. These radial zones are chosen such that for an observer at infinity they will contribute an equal amount of flux. The returning spectra (dashed lines) reveal a strong broadening of the reflection features. Moreover, as expected, the reflection spectrum also strongly depends on the ionization of the accretion disk. While for a larger ionization the spectrum roughly follows its incident spectrum, for a neutral disk strong absorption below 10\,keV greatly reduces the flux. These results already show that for a larger ionization we expect a stronger returning radiation flux in the relevant energy band probed by X-ray detectors. We note that at energies above 10\,keV the returning radiation spectra presented here are slightly affected by the limited modeling of the reflection hump in \xillver, which does not yet fully take into account the effects of the angular dependency of the Compton scattering cross section, but will soon be updated \citep{garcia2020}.

The full incident spectrum on each radial zone (thick line) is the combination of direct and returning spectrum. In case of an ionized disk we can see that the whole energy band is similarly affected, while for a neutral disk the returning spectrum greatly drops in flux below 10\,keV and is only a minor contribution to the total incident spectrum.

\subsection{New Relativistic Reflection Model Including Returning Radiation}
\label{sec:new-relat-refl-model}

In order to allow the study of relativistic reflection including the effect of returning radiation, the effect has to be included in a model that can be fit to observational data. We do so in the following, building upon the heritage of the \relxill framework. 

As described in the previous section, returning radiation results in an additional contribution to the irradiating flux that needs to be taken into account when calculating the total reflected spectrum emitted from the accretion disk. The returning spectrum, however, has a complex shape, which depends on the radius \rinc where it irradiates the disk (see Fig.~\ref{fig:relat_reflect_returnrad}). To correctly calculate the reflection produced by this returning spectrum would require a full radiative transfer reflection calculation \citep{Garcia2013a}. In principle, the results of such computations could be stored in a table (as is done in \xillver). However, this approach would significantly increase the dimensions of the \xillver reflection table by the parameters describing the returning radiation ($\rinc$, $a$, $R_\mathrm{in}$, $R_\mathrm{out}$) and therefore increase the number of spectra that need to be calculated and stored by at least a factor of 625 if only 5 grid points per parameter are calculated. Additionally, the time to interpolate these pre-calculated spectra for each model evaluation would also dramatically increase. It is therefore currently not feasible to perform these calculations. 

The only practical way to include returning radiation in a reflection model is to describe it as a contribution to the emissivity profile. As will be shown below, this approach correctly includes the main effect returning radiation has on relativistic reflection spectra. To describe the effect by a combined emissivity profile requires that the total irradiating flux, i.e., direct and returning, has a power law spectral shape at each radial zone \rinc, with the same index $\Gamma$ (Fig.~\ref{fig:relat_reflect_returnrad}). The incident flux is then given by
\begin{equation} \label{eq:nret}
  \Firrad = \Fsource+\Fret \approx \Ne(\Gamma, \Einc) \left( \varepsilon_p(\rinc) + C_F(\rinc)\varepsilon_o(\rinc) \right) 
\end{equation}
The emissivity of the returning radiation, $\varepsilon_o$, is given by Eq.~\ref{eq:emis-rrad}, which is weighted by a flux correction factor, $C_F(\rinc)$, which takes values between 0.3--1.2\footnote{For large values of $\Gamma$ and ionization $C_F$ can exceed 1. The reason is that a lower boundary of 0.1\,keV is used for the \xillver calculation and therefore also the energy flux calculation. While the energy band $>0.1\,$keV captures all relevant flux for the X-ray band, any effects below this boundary are not included.}. This factor ensures that while this total incident spectrum deviates from the original shape, it still contains the same energy flux as correctly predicted by the \xillver reflection calculations. Therefore
$C_F(\rinc)$ explicitly depends on the parameters of the accretion disk (density, ionization, iron abundance). Additionally, the flux boost factor $g^\Gamma$ is adapted to correctly take the change in flux due to the energy shift of a reflection spectrum instead of a power law into account. Therefore, similar to $C_F$, the flux boost of the returning radiation also depends on parameters such as the ionization of the accretion disk. More information on the dependence of both on $\Gamma$ and the ionization are given in Appendix~\ref{sec:flux-corr-fact}. Note that for a large ionization ($\log(\xi/\mathrm{erg\,cm\,s}^{-1}) \approx 3$) and $\Gamma>2$ both factors amount to roughly unity, meaning the emissivity profiles shown in the previous section (see Fig.~\ref{fig:pl_emis_profile_total}), can be regarded as representative for this combination of parameters.  

Re-visiting Fig.~\ref{fig:relat_reflect_returnrad} from the previous section shows the power law approximation of the returning reflection spectrum (dashed-dotted line) in comparison to the correctly smeared total irradiating spectrum. Generally, the correctly calculated irradiating spectrum (solid line) has only very mild and broad features and therefore no detailed reflection physics will be neglected by using the power-law approximation instead. The overall ionization structure will also be similar in both cases, as we set the energy flux of the irradiation to be the same. Also note that any small differences arising in the approximation will be completely smeared out, due to strong relativistic broadening of the reflection on the way to the observer.

Overall, this approximation fits well for large ionization, while there are larger deviations for the case of low ionization. This behavior can be readily understood looking at standard non-relativistic reflection spectra, which converge towards their incident spectra for very large degrees of ionization, but have a strong absorption below 10\,keV in the case of lower ionization. For this reason the ionization has the largest influence of all accretion disk parameters on the contribution of the returning radiation ($C_F=0.3$ for a neutral disk, slightly exceeding 1 for  $\log(\xi/\mathrm{erg\,cm\,s}^{-1}) > 3$). For lower ionization, the photon index of the primary radiation,  $\Gamma$, also influences the amount of reflected flux in a way that it is increased for small values of $\Gamma$ (see Appendix~\ref{sec:flux-corr-fact} for further details). 

Having described the total incident flux by the same spectrum with one
combined emissivity allows us to readily input this in the \relxill
model framework \citep[see, e.g.,][]{Dauser2010a}. A table is used to
store the information $(\Tf, g)$ of the ray-tracing simulations for
calculating the emissivity profile of the returning radiation. An
advantage of including the returning radiation as a contribution to
the emissivity profile is that it can be readily applied to any flavor
of the \relxill\ model, meaning it can be included for the lamp post geometry as well as for an empirical power law emissivity. Following our previous approach  with \relxill, the local
model including the returning radiation is publicly available in a
form to be used directly in all common X-ray data analysis
packages\footnote{The model is available
  \mbox{https://www.sternwarte.uni-erlangen.de/research/relxill/}. Returning
  radiation is included since version 1.5.}.

\section{Results: The effect of returning radiation in relativistic
  reflection modeling} \label{sec:results}

In the following we present the results when including returning
radiation in relativistic reflection modeling. As the effect of returning radiation is strongest for a compact corona (see Fig.~\ref{fig:emis_profile_zones}), it is likely that the inner disk is highly ionized due to the strong direct irradiation. We therefore assume
a larger ionization at the inner edge of the disk. For  easier interpretation and reproducibility we set the flux correction to $C_F=1$, treating the disk as a perfect reflector\footnote{A disk with   $\log(\xi/\mathrm{erg\,cm\,s}^{-1})=3$ and standard parameters would have $C_F=0.8$ (see   Appendix~\ref{sec:flux-corr-fact}).}. Furthermore, we will use the lamp post geometry as a simple but effective way of describing emissivity profiles similar to the ones obtained from measured spectra. This choice also allows for comparison with previous results obtained from modeling observational data without including returning radiation. 

\subsection{Effect of the Returning Radiation on the Observed Flux}
\label{sec:effect-return-radi}

The most basic observable effect of the returning radiation is the increase of reflected flux in the observed spectrum.
\begin{figure}
  \centering
  \includegraphics[width=\columnwidth]{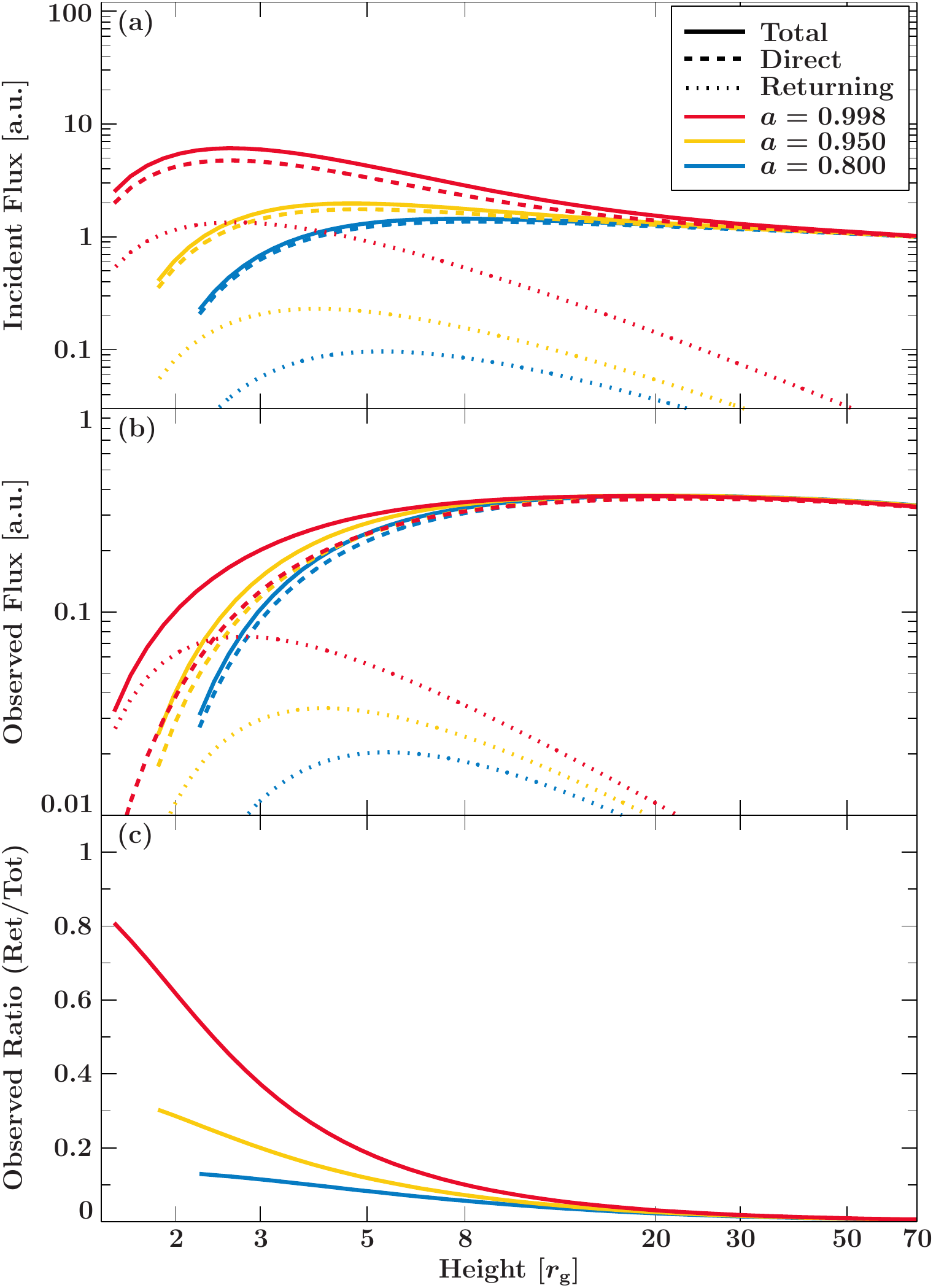}
  \caption{Total photon flux irradiating the disk for a primary power law with
  $\Gamma=2$. Dashed lines show the contribution of the primary radiation, dotted lines indicate the contribution of the returning radiation. (a) Incident photon flux in the frame of the disk. (b)  Same for the observed, assuming an inclination of $\incl=45^\circ$ and a perfect reflector ($C_F=1$). (c) Ratio of the contribution of returning radiation to the total observed flux.   }
  \label{fig:return_flux_fraction}
\end{figure}
Figure~\ref{fig:return_flux_fraction}a shows the total flux irradiating the disk from the primary and the returning radiation for different primary source heights at fixed emitted flux in the source frame. It can be seen that returning radiation (dotted lines) produces the strongest contribution for a low source height. Overall the returning flux integrated over the disk makes up, at most, 10\% of the total flux irradiating the disk. As expected, this increase in flux is largest for maximal spin ($a=0.998$) and quickly diminishes to a negligible amount for smaller spin values. This shows that even for a very compact source at $h=2\rg$ the fraction of returning flux compared to the total incident flux is small, while we have seen from the emissivity profile (see Fig.~\ref{fig:emis_profile_zones}) that the returning radiation dominates the irradiation of the disk at $r>4\rg$. The reason is the extremely focused irradiation of the very inner accretion disk by such a compact corona.

More interesting is the effect of the returning radiation on the observed flux (Fig.~\ref{fig:return_flux_fraction}b). In the case of a compact source $h<3\rg$ and extreme spin ($a=0.998$), returning radiation makes up 40-80\% of the total observed flux, meaning it can be up to a factor 4 stronger than the direct reflection. While this sounds contradictory at first from what we have seen for the irradiating flux (Fig~\ref{fig:return_flux_fraction}a), it can be easily understood when considering the strong gravitational redshift the photons reflected at the very inner disk exhibit before being observed. By returning to the outer disk the reflected photons can therefore partly avoid the flux reduction due to the gravitational redshift as, due to the fast rotation of the accretion
disk, a significant fraction of the photons emitted at the inner disk also see special relativistic blueshift. In this way, the strong flux decrease of radiation emitted at the inner disk is mitigated by a secondary reflection at larger radii. Without returning radiation (dashed line in Fig.~\ref{fig:return_flux_fraction}b), reflection from a primary source located at a small height is strongly suppressed in flux due to the strong gravitational redshift experienced by photons observed from these regions. In concrete numbers this means that while without returning radiation the observed flux from a compact corona at $h=2\rg$ is reduced by a factor 10, it is only reduced by a factor 2.5 when including returning radiation.

For black holes with large spin, the secondary reflection therefore dominates the observed reflection for $h<3\rg$, as it recovers most of the flux lost due to gravitational redshift of the direct radiation. This effect, however, depends strongly on the spin of the black hole. As is evident from the ratio between the flux of observed reflection induced by returning radiation with respect to the flux due to total reflection (Fig.~\ref{fig:return_flux_fraction}c), for a spin of $a=0.95$ returning radiation can contribute at most 30\% to the total flux and for $a=0.8$ the contribution drops to 10\% and is strongly decreasing with an increasing primary source height. Even for the largest spin values, the returning flux will only play a minor role for heights above $10\rg$.

In conclusion, the observable effect of the returning radiation strongly depends on the spin and on the height of the primary source. It ranges from being the dominant contribution to the observed relativistic reflection for extreme configurations of spin and very small corona height, to being negligible for spins $a\lesssim 0.8$ and heights $\gtrsim 10\rg$.

\subsection{Line Profiles and Relativistic Reflection Spectra}
\label{sec:relat-refl-spec}

\begin{figure}
  \centering
  \includegraphics[width=\columnwidth]{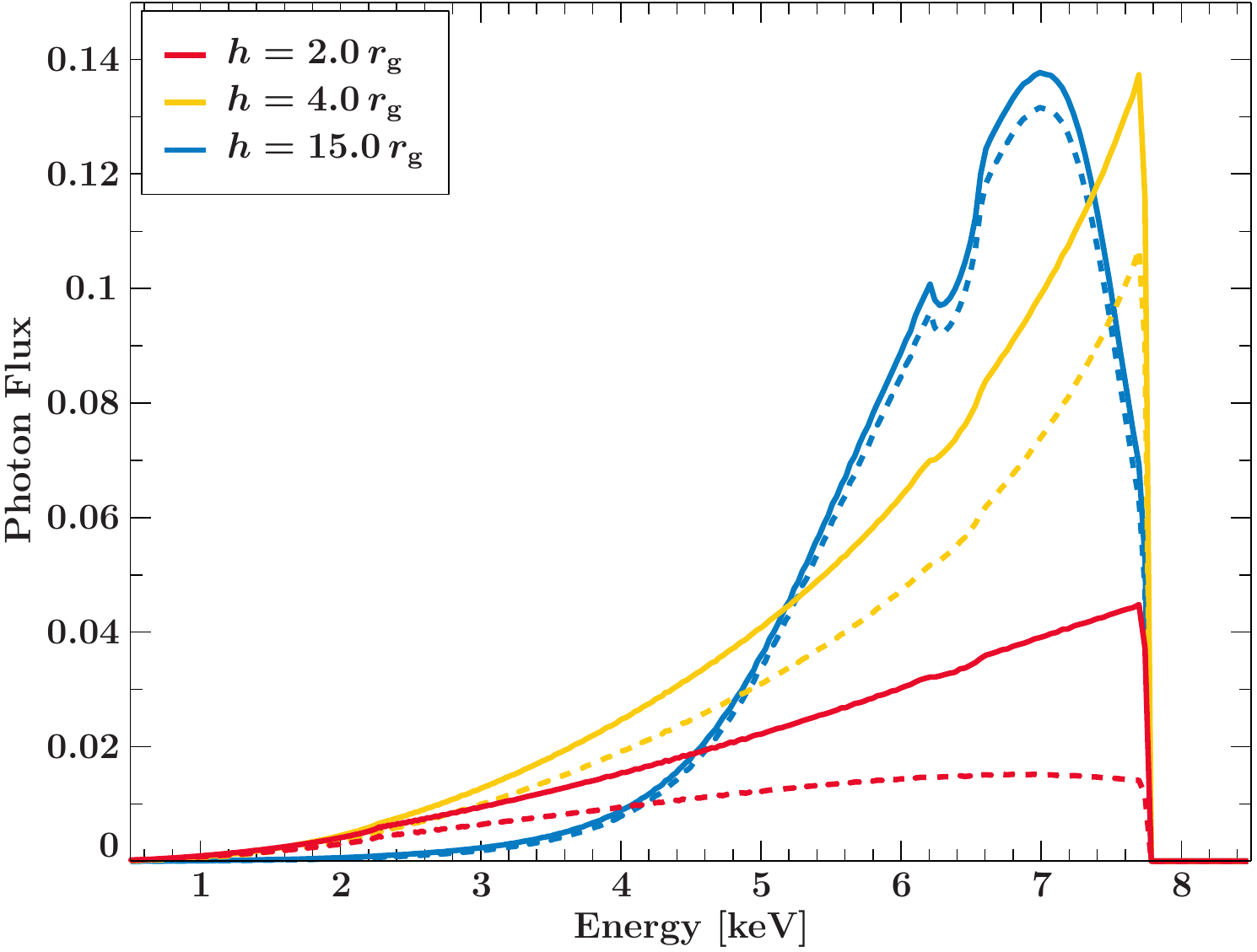}
  \caption{(Left) Line profiles for the lamp post geometry ($a=0.998$, $\incl=60^\circ$) and an irradiating power law with $\Gamma=2$. Solid lines  include the returning radiation, dashed shows lines excluding returning radiation.}
  \label{fig:lines_lp}
\end{figure}
Line profiles including returning radiation for different source
heights are shown in Fig.~\ref{fig:lines_lp}. The shape of the lines
does not differ in a significant way from profiles that do not include
returning radiation, however their flux is enhanced. From over 50\% increase in flux for a small source height ($h=2\rg$), it decreases for larger values of height, in agreement with the results discussed in Sect.~\ref{sec:effect-return-radi}. This behavior has an effect on the reconstruction of the source height from data. For example, when including returning radiation, the line shape predicted for a source with height of $2\rg$ is very similar to that predicted for a source with height of $3\rg$ when ignoring returning radiation. The reason for this similarity between line profiles can be understood from Fig.~\ref{fig:pl_emis_profile_total}, as the originally very steep emissivity profiles for $h=2\rg$ are flattened and more similar due to the returning radiation on the outer disk.

\begin{figure}
  \centering
  \includegraphics[width=\columnwidth]{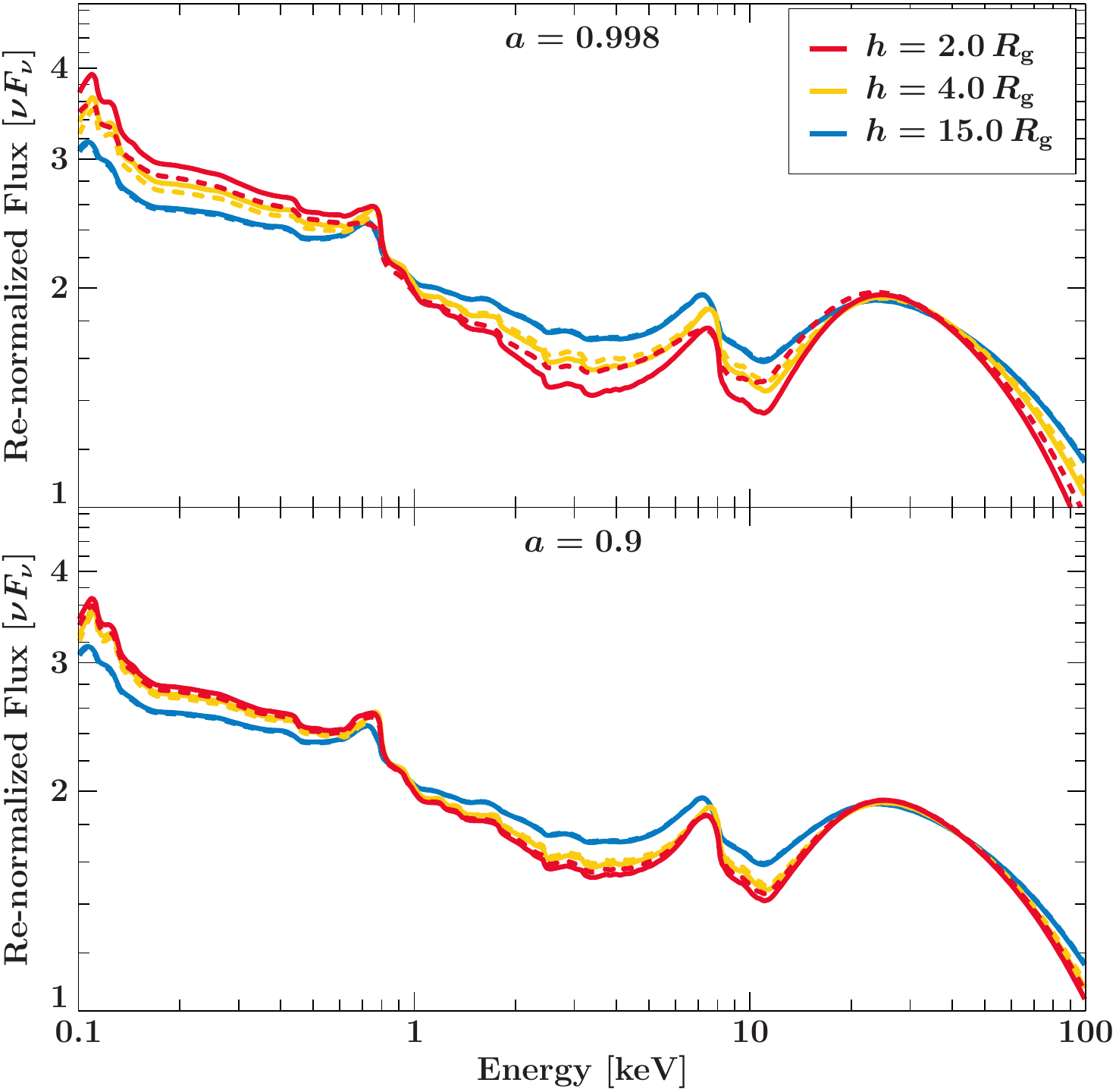}
  \caption{Relativistic reflection spectra calculated with
    \texttt{relxilllp} for different heights of the primary source
    with fixed intrinsic luminosity. The spectra include the direct
    radiation and are re-normalized to the same $\nu F_\nu$ flux, in
    order to highlight the differences in spectral shape. Solid lines
    depict the new model including returning radiation, dashed lines
    exclude returning radiation. (Upper panel) spin set to $a=0.998$.
    (Lower Panel) $a=0.9$. In both cases, the power law index of the
    irradiating photons was set to $\Gamma=2$, the ionization to $\log(\xi/\mathrm{erg\,cm\,s}^{-1}) =  3.1$, inclination $\incl=60^\circ$, and the iron abundance
    set to  solar. }
  \label{fig:relxill_spectra}
\end{figure}
The full relativistic reflection spectra are shown in Fig.~\ref{fig:relxill_spectra}. These spectra include the direct radiation from the source, as well as reflection from direct illumination and returning radiation. It therefore represents the total reflection spectra observed for a lamp post primary source. It can be seen that including returning radiation (solid lines) leads to stronger and narrower reflection features compared to models neglecting returning radiation (dashed lines). While for an extreme spin  of $a=0.998$ large differences are visible when including returning radiation,  for $a=0.9$ these differences make up only a small contribution to the overall spectrum.

\subsection{Returning Radiation Boosting the Reflection Strength}
\label{sec:return-radi-boost}

Observationally, the reflection fraction $R_\mathrm{F}$ is an important measure of the geometry of the accreting system \citep{Dauser2014a}. This quantity is defined as the ratio of photons emitted from the primary (lamp post) source that will hit the accretion disk compared to those which directly propagate to the observer. Meaning, it is a measure of the fraction of primary photons that will produce reflection and therefore characterizes the angular emission profile of the primary source. For a lamp post source at low height, it will be mainly influenced by strong light-bending, which focuses most photons towards the accretion disk.

For a low source height and large spin, the reflection fraction can reach values of 10, that is, the majority of the emitted flux will intercept the disk and produce reflection \citep{Dauser2016b}. As the reflection fraction is purely defined by the primary source, however, the predicted reflection fraction is not affected by the returning radiation and is therefore solely a measure of the emission characteristic and geometry of the primary source.

The \emph{reflection strength} $R_\mathrm{S}$, on the other hand, is a measure of the observed reflected flux with respect to the observed flux directly emitted by the primary source. To be consistent with \citet{Dauser2016b}, we define the reflection strength as the ratio of the observed energy flux of the reflection with respect to the direct flux from the primary source in the energy band of 20--40\,keV. The reason for choosing this band is that electron scattering dominates the reflection in this band, leading to the Compton hump, making $R_\mathrm{S}$ largely independent of parameters such as the ionization or iron abundance. While a large reflection fraction typically leads to a large reflection strength, i.e., a reflection dominated spectrum, the exact value also strongly depends on the inclination of the system \citep{Dauser2016b}.

\begin{figure}
  \centering
  \includegraphics[width=\columnwidth]{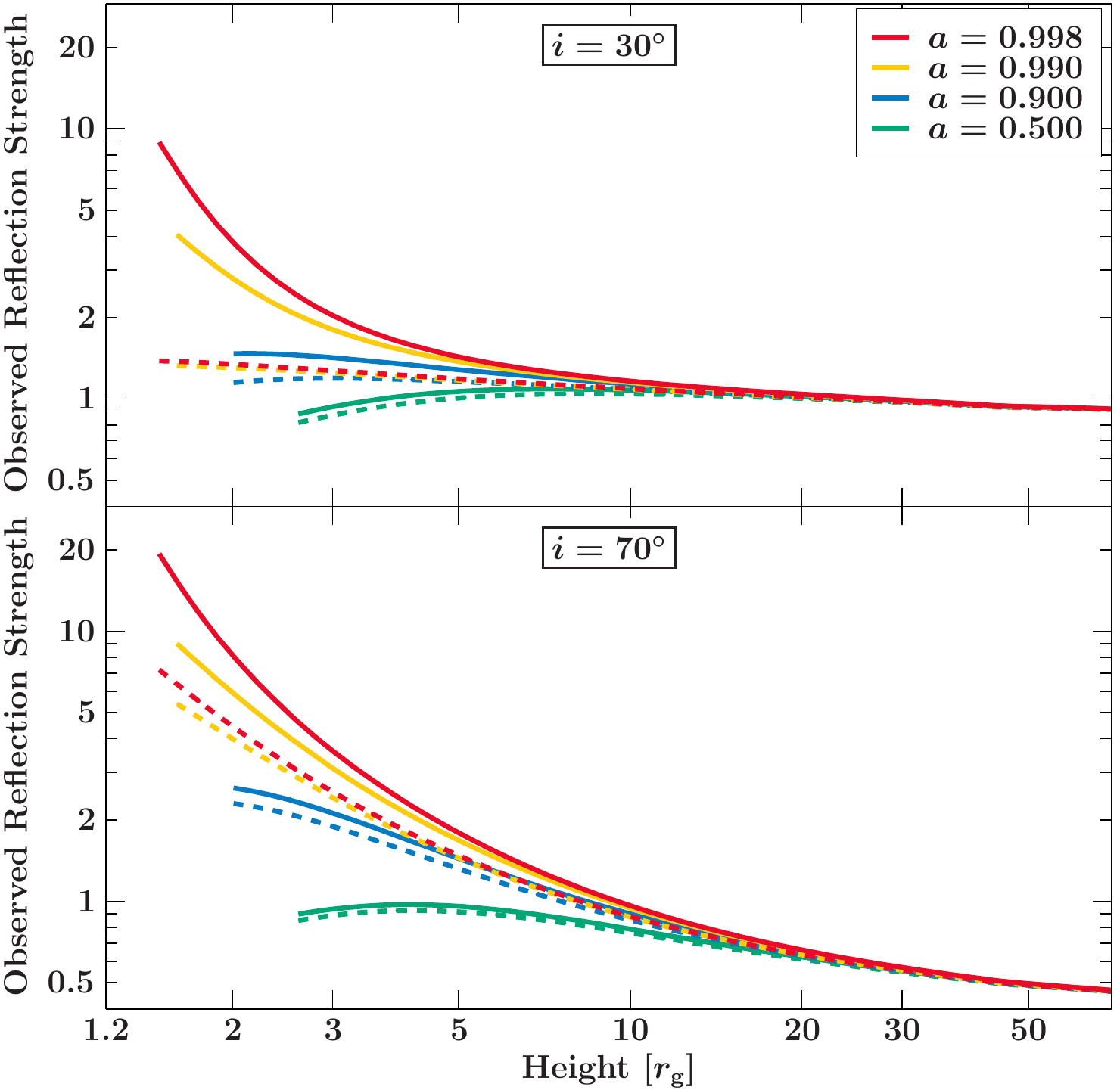}
  \caption{Reflection strength for spectra observed at different
    observed  inclinations $\incl$. As in previous figures, solid
    lines show  models including returning radiation while dashed
    lines do not include returning radiation. }
  \label{fig:reflection_strength}
\end{figure}
Figure~\ref{fig:reflection_strength} shows the reflection strength $R_\mathrm{S}$ including returning radiation as a function of primary source height (solid lines) in comparison to only direct reflection (dashed lines), for different values of spin and inclination. Generally, as expected the returning radiation boosts the reflection strength. For the largest spin and low source heights, a reflection strength of above 10 is easily possible. This result implies that the observed spectrum would be fully reflection dominated. The reflection strength quickly drops to below 2 for primary source heights between $h = 3\rg$-$5\rg$. As expected, for spins $a<0.9$ returning radiation barely influences $R_\mathrm{S}$ and no large values of the reflection strength are predicted either. This underlines the conclusion from \citet{Dauser2016b} that a reflection dominated spectrum can only be predicted by the lamp post configuration for very high values of black hole spin.

While returning radiation generally strongly boosts the reflection strength, Fig.~\ref{fig:reflection_strength} also shows that this boost is stronger for low inclination. When only taking direct reflection into account, those low inclination systems never show a large reflection strength. However, returning radiation can boost this up to values of 10. This means that including returning radiation decreases the variation of $R_\mathrm{S}$ with respect to the inclination to the system and therefore regardless of the inclination a large reflection strength is predicted for low height sources and rapid black hole rotation. This can be understood as the outer part of the accretion disk now plays a larger role and therefore the slower particle motion in the disk implies that the observed flux depends less on the inclination towards the system.

\subsection{Observational Bias of Neglecting Returning Radiation}
\label{sec:simulations}

We established in the previous sections that for the arguably most interesting parameter space of low source height and high black hole spin, the returning radiation significantly influences the relativistic reflection. Therefore a major question is the systematic bias contained in the black hole and reflection parameters that were obtained with reflection models that ignored returning radiation.

In order to answer this question we simulate realistic combined \textsl{XMM-Newton} and \textsl{NuSTAR} observations with the new \relxill\ model that includes returning radiation, and then model these with a reflection model that does not include returning radiation. Specifically, we simulate an observation of a typical bright Seyfert~1 galaxy with a black hole with $a=0.998$ seen under an inclination of $30^\circ$, a total flux of $1\times10^{-10}\,\mathrm{erg}\,\mathrm{cm}^{-2}\,\mathrm{s}^{-1}$ in the 0.5--10\,keV band, and 100\,ks of total effective exposure in both missions. The primary continuum is a power law continuum with $\Gamma=2$. With $\log(\xi/\mathrm{erg\,cm\,s}^{-1})=3.1$ the accretion disk is strongly ionized. We assume that the disk has an iron abundance of $A_\mathrm{Fe}=2$ with respect to solar abundances \citep[using][abundances]{Wilms2000a} and a Galactic absorption of $N_\mathrm{H}=4\times10^{20}\,\mathrm{cm}^{-2}$. The simulation is then performed for different values of primary source height $h$.

\begin{figure}
  \centering
  \includegraphics[width=\columnwidth]{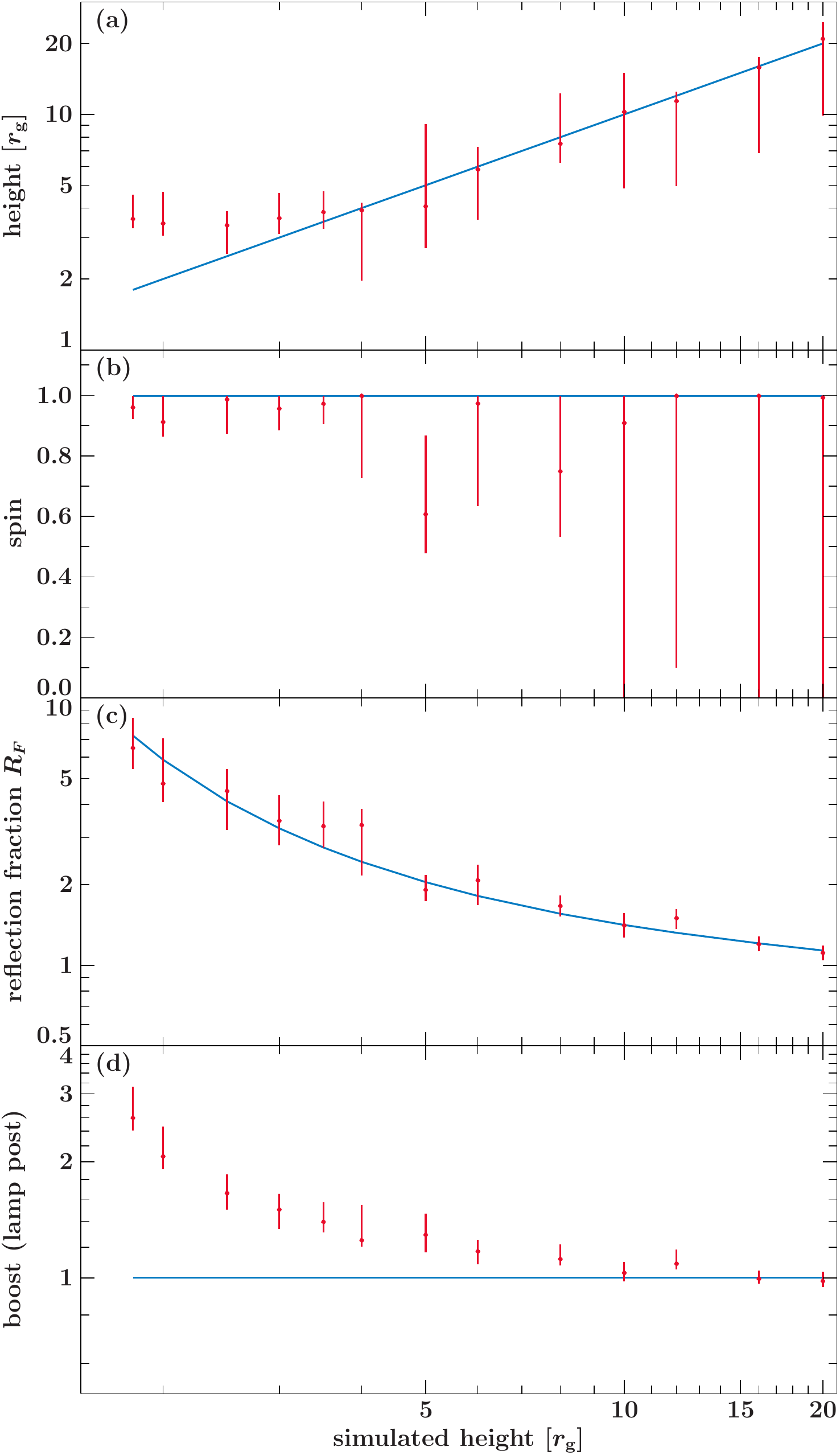}
  \caption{Results from simulating with the new \relxill model and fitting this data with the normal \relxill\ model without returning radiation (red data points) for different values of the height of the primary source. The spin was set to $a=0.998$, $\Gamma=2$, $\log(\xi/\mathrm{erg\,cm\,s}^{-1})=3.1$, and the iron abundance  $A_\mathrm{Fe}=2$. The blue curves show the input values used for the simulation. Uncertainties are given at 90\% confidence.} 
  \label{fig:simulation}
\end{figure}

Figure~\ref{fig:simulation} shows the result of fitting the simulated data with a \relxill\ model that does not include returning radiation. Down to a primary source height of $h\approx4\rg$ the best fit results reproduce the input values (Fig.~\ref{fig:simulation}a). This is expected, as in these cases the returning radiation does not play a dominant role in the relativistic reflection spectrum. For lower heights, however, the height fitted with the old model is always around 3-4\rg, even for a very small true height of $h=1.8\rg$. These results reflect the findings from the previous sections that the returning radiation flattens the emissivity profile such that reflection spectra for low coronal heights are practically indistinguishable. Measurements of the spin are not affected by the returning radiation (Fig.~\ref{fig:simulation}b). Similarly, other parameters such as the iron abundance are also properly recovered. Overall, we note that all fits with the standard \textsc{relxilllp} model are very well able to explain the spectrum including returning radiation in all cases.

Figure~\ref{fig:simulation}c shows that the reflection fraction is also well recovered. Taking into account that the reflection fraction describes the intrinsic irradiation of the disk, this is expected from Fig.~\ref{fig:return_flux_fraction} where the additional flux impinging on the accretion disk due to returning radiation is at most 10\%. In principle, the increased reflection strength could indeed be compensated by a larger reflection fraction. However, in our simulated high quality spectra a larger height of the primary source results in a better fit and fully compensates for the flatter emissivity profiles predicted for very low source heights ($h<4\rg$) when including returning radiation. 

However, for this case of a very compact corona (i.e., $h<4\rg$), we obtain a larger height in our fits, meaning that the predicted reflection fraction in this case is decreased. Therefore, while the reflection fraction remains constant, compared with the increased height it is now over-predicting the reflection contribution. This is best visualized by looking at the boost-parameter $\mathcal{B}$, which is defined as the ratio of the fitted reflection fraction with respect to the predicted reflection fraction. For an ideal lamp post corona source, we would have $\mathcal{B}=1$, as assumed in our simulations. It can be seen in Fig.~\ref{fig:simulation}d that for a low height primary source we obtain a boost parameter significantly larger than 1. While smaller values of $\mathcal{B}<1$ can for example be explained by an outwards moving corona \citep[see, e.g.,]{Dauser2013a}, values of $\mathcal{B}>1$ are not easily explained as the lamp post configuration already gives the strongest reflection compared to other coronal geometries and a rapidly inward-moving corona to artificially increase $\mathcal{B}$ defies any physical reasoning. 

Therefore the larger height in the fits compensates for the flattening of the emissivity profile. As for this configuration a lower reflection strength is predicted, the boost parameter and thus also the reflection fraction is increased. This means that without including returning radiation, the results obtained will lead to an over-prediction of the reflection fraction for the fitted height and therefore seemingly too strong irradiation of the accretion disk.

\section{Discussion} \label{sec:discussion}

\subsection{Comparison to previous results } 

In the following we will compare our results with previous work on
returning radiation. In their study of dissipation in a Novikov-Thorne disk, \citet{reynolds2004} found that the radial dependence of the flux from a corona, i.e., the emissivity, flattens when returning radiation is included, while overall the total emissivity is increased. Our results are in perfect agreement with these earlier findings, when taking into account that we are assuming a lamp post geometry. This comparison therefore shows that the overall effect of returning radiation is to add flux at larger radii and to flatten the otherwise steep emissivity profiles, regardless of the assumed geometry of the corona (see Fig.~\ref{fig:emis_profile_zones}, which shows only the additional irradiating flux created  by returning radiation). Similarly, comparing our results to the lamp post configuration presented by \citet{niedzwiecki2008}, we are in good agreement on how returning radiation influences the broad line shape and also with the result that returning radiation dominates the observed reflected flux for a very compact corona and high spin. 

In general, our results also compare well with studies for a neutral accretion disk \citep{niedzwiecki2016,riaz2021}, where the main result is an increase in reflected flux for energies $>$10\,keV. While we agree that in this case these energies are more affected by returning radiation, we find that for a low lamp post height the broad iron K$\alpha$ line is also boosted by returning radiation. We note, however, that such a neutral accretion disk is not expected at the inner accretion disk and also typically not found in observations. Therefore the effect of returning radiation is most relevant for spectra from ionized reflection, where the influence of the returning radiation is equally present at softer X-ray energies (see Sect.~\ref{sec:relat-refl-spec}).

In a different approach, \citet{wilkins2020} determined the effect of returning radiation on the observed relativistic reflection spectrum for one parameter combination ($h=5\rg$, $A_\mathrm{Fe}=8$, low disk ionization) and variable photon index. While in this case the rest-frame reflection itself was modeled in great detail, the energy shift of all incident returning radiation at a certain radius \rinc on the disk was averaged and only a single value used to shift the spectrum. As we have shown in Fig.~\ref{fig:gshift_line}, however, each ring on the disk will see a large range of energy shifts. This effect results in strong broadening of the returning radiation, leading to a very smooth returning spectrum without any narrow line features. Similarly, \citet{ross2002} also used  a non-broadened reflection spectrum as input in their modeling. While their result that secondary reflection increases the strength of the iron line agrees with our findings, it is not applicable in the relativistic regime we consider, where the iron line is strongly broadened when returning to the disk. The energy shift also strongly influences the flux boost or reduction from certain parts of the disk and is therefore an essential part in determining the overall effect of the returning radiation. This crucial difference to earlier works explains why the effect of the returning radiation predicted here is much larger than previously claimed. We note that the underlying ray-tracing results are in good agreement with our results (see Appendix~\ref{sec:comparison-with-ak00-and-wilkins}).

\subsection{Angular Dependency of the Returning Radiation}
\label{sec:angul-depend-return}

The irradiation of the disk due to returning radiation has a different angular distribution to that due to the corona. While a full calculation of these effects goes beyond this paper, in the following we will briefly discuss the potential effects. As shown in Fig.~\ref{fig:dist-single-rays}, the returning radiation will mostly be striking the disk at a shallower angle compared to the direct radiation. This will change the shape and flux of the reflected spectrum. For example, the increased flux at soft X-rays will be changing the ionization balance of the gas \citep[see Fig.~5 of][]{Dauser2013a}. A similar spectral shape will be obtained by effectively increasing the ionization  \citep[by up to a factor of five, see][]{Dauser2013a}.\footnote{We note that current detailed reflection models such as \textsc{reflionx} \citep{Ross2005a} or \xillver \citep{Garcia2013a} assume $\delta = 45^\circ$ as incident angle of the radiation.}
 
Given that many of the reflected photons have now been scattered through right angles, the associated Compton reflection will also be more highly polarized. Therefore we conclude that returning radiation will lead to a higher degree of polarization of the reflected spectrum. These results are in agreement with \citet[][]{schnittman2009}, who find that the grazing incidence of returning thermal emission from the disk dominates the polarization signal at higher energies. Further investigation is needed, however, to determine how large this effect would be and if it is detectable by future X-ray missions such as \textsl{eXTP} \citep{zhang2016} or the recently launched \textsl{IXPE} \citep{weisskopf2016}.

In the hard state of Cyg~X-1, \citet{chauvin2018} find a comparably low polarization degree in the Compton hump dominated band, and therefore argue that the spectrum can not be reflection dominated. As significant returning radiation would only increase the polarization degree, this result directly implies that returning radiation can not be important in the hard state of Cyg~X-1. Note that in their spectral-timing analysis of Cyg~X-1, \citet{mastroserio2019} find $h~\sim 9\rg$, which is consistent with our results that returning radiation is not significant at such a height of the primary source.

\subsection{Importance for Reverberation Measurements}

The inclusion of returning radiation will also have a significant impact on the predicted reverberation signal in spectro-timing studies, since returning photons follow a longer path on their way to the observer than primary photons. The increase in path length can be large, since many returning photons are reflected for the second time at a much larger disc radius than the first. Returning radiation may in fact be more important for the creation of reverberation lags than it is for the spectrum, since we find that it only becomes important for the spectrum for source height $h \lesssim 3r_g$ (Fig.~\ref{fig:simulation}) whereas \citet{wilkins2020} found that returning radiation increases the time lags in the iron K-band by $\sim 50\%$ even for $h=5\rg$. Moreover, the simplified treatment of the energy shift employed by \citet{wilkins2020} could even mean that these authors underestimated the importance of the effect on the time lags. We also note that, whereas we find that higher order reflections can be ignored in the calculation of the spectral shape, this is not necessarily the case for the time lags. This is because even though the flux of each order is less than that of the previous order, the time lag is longer.

The inclusion of returning radiation may solve some long standing problems in the literature: the iron K feature in the lag spectrum often appears to be too strong compared to the Fe K$\alpha$ flux \citep{Mastroserio2020,Zoghbi2020}, and the source height inferred from timing alone is greater than that inferred from the spectrum alone \citep{wang2021}. Returning radiation, however, will increase the time lags, leading to a larger than expected iron feature in the lag spectrum. Therefore fitting a model to the lag spectrum that neglects returning radiation may greatly over-estimate the source height, whilst comparing the same model to the energy spectrum alone will only slightly over-estimate $h$.

\subsection{Reflection Dominated Spectra}

In the low flux state of bare Seyfert~1, the reflection has often been found to dominate the observed spectra (see, e.g., 1H0707$-$495, \citealp{Fabian2012a} or 1H0419$-$577, \citealp{jiang2019}). Using full relativistic reflection models such as \relxill or \textsc{reltrans}, the measured reflection fraction can be compared to the expected one. The ratio of those two values is called the boost parameter $\mathcal{B}$. Especially in AGN such as 1H0707$-$495 \citep{Fabian2012a,kara2015,boller2021}, but also in the Galactic black hole binary MAXI~J1820$+$070 \citep{wang2021,you2021} this boost parameter was found to be $\mathcal{B}>1$, meaning that the reflection fraction was under-predicted by the lamp post model. No viable interpretation or alternative explanation has been given to explain these very reflection dominated spectra. As shown in our simulations (see Fig.~\ref{fig:simulation}), for those cases of a compact corona, not taking into account returning radiation could lead to an artificial increase of this boost parameter. This highlights the importance of including returning radiation, potentially resulting in previous measurements of very low source heights being more consistent with the lamp post model.

Including returning radiation in the spectral modeling, these reflection dominated spectra are therefore easily explained. For geometries with high spin and compact corona, which are typically found in those reflection dominated sources, returning radiation will significantly increase the observed reflection strength by a factor of a few (see Fig.~\ref{fig:reflection_strength} and Sect.~\ref{sec:return-radi-boost}). Due to the very strong dependence of the reflection strength on returning radiation, we expect that in the observations mentioned above the new \relxill model will be better able to consistently explain the data within the lamp post geometry. Even an out-flowing corona, which decreases the reflection strength \citep[see][]{Dauser2013a}, could lead to strong observed reflection because of returning radiation.

The enhanced reflection from the outer disk due to returning radiation might also be able to explain the rather unphysically strong and ionized distant reflection component that is often required for X-ray binary black holes \citep[see, e.g.,][]{shreeram2020}. Our results in Sect.~\ref{sec:relat-refl-spec} show that including returning radiation leads to a stronger and narrower Fe~K$\alpha$ reflection feature (see Fig.~\ref{fig:relxill_spectra}). It is therefore possible that this stronger and narrower feature was at least partly wrongly identified as ionized distant reflection. The inclusion of returning radiation in reflection models would somewhat subsume the distant reflection component, which may then be weaker and less ionized than found with previous models. A proper assessment of whether or not returning radiation can partially explain the unphysical ionized distant reflector requires detailed data analysis, which is beyond the scope of this paper.

\subsection{The problem of the steep emissivity}
\label{sec:steep-emissivity}

A large number of observational studies of accreting black hole systems require the coronal source height to be very small $h<4\rg$ \citep[e.g.,][]{Dauser2012a,Parker2014a,kara2015,walton2020,caballero-garcia2020,boller2021}.
In a few cases, extremely compact source geometries of $h<2\rg$ were found \citep[e.g.,][]{Fabian2012a,beuchert2017}. While these extreme results were already hard to explain \citep{niedzwiecki2016,ursini2020}, the additional bias by not taking the returning radiation into account would even further reduce the inferred coronal height  (see Sect.~\ref{sec:simulations}). Therefore the true source height would be even closer to the black hole than predicted from previous relativistic reflection models.  However, as shown in Sect.~\ref{sec:simulations} it should have been impossible to obtain measured heights smaller than even 3\rg with the standard assumptions on the disk.

The reason for these contradictory results can be best understood when considering the emissivity profile. The above results of a very compact corona imply a steep emissivity profile at the inner disk, significantly deviating from the standard $r^{-3}$ profile \citep{Dauser2013a}. These steep profiles are consistent with results from reflection models where empirical emissivity profiles are used \citep[e.g.,][]{wilms2001a,marinucci2014,jiang2018}, as well as with direct measurements of the emissivity profile \citep{Wilkins2011a,Parker2014a} in a multitude of sources. Therefore, the basic conclusion of these studies is that the majority of reflected flux  originates from the very innermost disk.

As we have emphasized in this paper before, returning radiation increases the irradiation of the outer disk. Returning radiation therefore makes it impossible for any geometrical configuration, including the lamp post geometry, to predict such a steep emissivity profile (see Sect.~\ref{subsec:emissivity-profile}), which is in contradiction to all above-mentioned observations. In fact, the steepest profile is not found for the highest spin, but for a medium-high value of 0.9, which is the sweet-spot between being still close enough to the black hole to have a steep primary emissivity profile, but not close enough to produce that much returning radiation to irradiate the outer parts of the disk and flatten the emissivity profile. Moreover, as shown in Fig.~\ref{fig:relxill_spectra}, the extreme broadening of the reflection features can not be produced in the presence of returning radiation.

However, following our current assumptions on the corona and disk, returning radiation inevitably has to be present for a high spin and compact source. Therefore the question is whether the very steep emissivity profile is a bias introduced due to physically incomplete modeling of the system, or whether it is due to erroneous assumptions on the primary source geometry. Concerning the latter point, we note that so far the only accretion geometry that can explain steep emissivity profiles is the lamp post. Moreover, as we have shown, even if only the inner edge of the accretion disk is irradiated by the corona, which results in the steepest possible profile, returning radiation from the inner edge will significantly flatten the profile and dominate the disk irradiation.

There are two simplifications we have made that could influence the emissivity profile or alter the effect of the returning radiation on the emissivity. Firstly, we have assumed a geometrically thin disk. Using a disk with a finite thickness will certainly influence the flux of the returning radiation. This scenario will depend on the mass accretion rate, the shape, density, and ionization of the inner disk, which can take a complex form \citep{marcel2018}. A simple thick disk with a constant $H/r$ ratio or an outer flared disk will not influence the inner disk emissivity directly.  Additionally, for such a thick disk a larger fraction of photons will be able to return to the outer disk and therefore increase the irradiation of the outer disk.  Relativistic reflection from a radiation dominated \citet{Shakura1973a} disk, which lead to a disk with constant $H$ tapering down to zero at the ISCO, was studied by \citet{taylor2018}, who concluded that such a configuration could block our view onto the inner regions for large inclination angles. In turn, such a shape of the inner disk could focus more returning radiation onto the inner parts, while at the same time shielding the outer regions from direct coronal emission \citep{taylor2018} and also from returning radiation. These effects will increase the emissivity observed from the inner disk. In the extreme case, such a disk shape could be comparable to that posited for Ultraluminous X-ray Sources, where a funnel is potentially created and radiation scattered (i.e., returning in our picture) multiple times \citep[e.g.,][]{dauser2017}. However, for typical reflection dominated sources the mass accretion rate is much lower than required for such a disk shape, and therefore this is an unlikely scenario for the objects where strong relativistic reflection is typically observed. More detailed studies that are beyond the scope of this paper are required to assess the effect of returning radiation for such radiation dominated disks.

Furthermore,  we have assumed a constant ionization throughout the disk. Because the irradiating flux is strongly variable with radius, however, we expect an ionization gradient on the disk, as the ionization is directly linked to the incident flux \citep{Dauser2013a,Garcia2014a}. Previous studies have also shown that an ionization gradient will lead to steeper emissivity profiles \citep{Svoboda2012a,kammoun2019a}. However, note that the ionization depends strongly on the density of the accretion disk as well, and therefore using, e.g., an $\alpha$-disk density profile leads to a very different gradient \citep[see, e.g.,][]{ingram2019,shreeram2020} than  the aforementioned publications, which assumed constant density disks. Note that since the irradiation is focused on the inner edge of the disk, the assumption on the torque at the ISCO will strongly influence the accretion disk properties such as the density and therefore also the ionization.

This behavior can also be understood in terms of the correction factor $C_F$ (see Fig.~\ref{fig:C_F}). It is defined as the fraction of locally emitted reflected radiation with respect to the irradiating flux in the X-ray energy band where the emissivity is measured. Increasing the ionization from $\log(\xi/\mathrm{erg\,cm\,s}^{-1})=2$ to $4$ will increase the reflected flux in the X-ray band by up to a factor of 3, which is in agreement with the above mentioned studies leading to a steeper observed emissivity for a disk with an ionization gradient.  Note that the bolometric emissivity will be constant, as a larger fraction of the radiation is thermalized for lower ionizations, which is outside the observed X-ray band.

\section{Summary and Conclusions}
\label{sec:summary-conclusions}

We presented calculations of returning radiation from relativistic reflection emitted from the accretion disk. The resulting additional irradiation was included in the \relxill reflection modeling framework,  allowing us to model relativistic reflection spectra including returning radiation for the first time. While we investigated disk irradiation and reflection by a lamp post corona, our results only depend on the incident spectrum and emissivity profile and are therefore applicable to all similar emissivity laws. Using this approach, we could identify the main behavior of the returning radiation:
\begin{itemize}
\item Returning radiation is primarily important for high spins of $a\gtrsim 0.9$ and compact coronae at heights $h< 5\rg$. We note that in the case that the primary radiation is not dominated by the corona, but thermal black body emission from the disk, returning radiation can also be detectable for low spin \citep{connors2020}.
\item The main effect of returning radiation is to flatten the emissivity profile, which is caused by primary reflected photons from the very inner disk irradiating the outer disk.
\item Due to strong energy shifts, the primary reflected spectrum returning to the accretion disk does not exhibit any narrow reflection features, but is strongly relativistically broadened.
\item Returning radiation enhances the amount of reflection. For extreme values of compactness ($h<2\rg$) and maximal spin ($a=0.998$), the additional reflection caused by returning radiation dominates the flux of the observed reflection spectrum.
\item Simulations of observable spectra show that previous studies that did not include returning radiation likely overestimated the coronal height, while simultaneously underestimating the reflection fraction in comparison with the predicted reflection fraction from the determined height.
\end{itemize}

Combining these general results with the current state of observational results obtained from relativistic reflection measurements, we can draw fundamental conclusions for these systems:
\begin{itemize}
\item The apparent discrepancy of the lamp post model under-predicting the reflection strength in some sources that show very strong reflection (e.g., 1H0707$-$495) can potentially be solved by accounting for the contribution of the returning radiation to the reflected flux.
\item Including the effect of returning radiation in reverberation timing studies will likely lead to an overall increase in the expected time lags, meaning the primary source height has been overestimated in previous studies that did not take returning radiation into account. This effect could explain the discrepancy that the source height found by modeling time lags \citep{alston2020,wang2021} is generally larger than found from spectral analyses of the same observation.
\item The very steep emissivity profiles and low source heights that are often required to describe the spectra and that are typically explained by a very compact corona are in contradiction with the flatter emissivity profile when including returning radiation. A possible explanation for this discrepancy could be an ionization gradient in the accretion disk, as a highly ionized disk will have an emissivity that is a factor 3-4 larger in the observed X-ray energy band than a more neutral disk.
\end{itemize}
In the last decade, detailed observational studies  analyzed the various parameters entering the relativistic reflection studies in order to determine the inner accretion geometry. Due to the complexity of those parameters, a more self-consistent approach is required to understand these systems. While typically adding more detailed physics leads to either an increase of the number of parameters or requires additional assumptions, including returning radiation comes at minimal cost to the modeling. Its effect can be directly added to any primary disk irradiation without prior assumptions. One of the biggest uncertainties is the density profile of the accretion disk. With the new generation of relativistic reflection models, such as \relxill, it will be possible to fit the density profile, while the remaining parameters such as the ionization of the disk and the effect of the returning radiation will be automatically and consistently taken into account.

\section*{Acknowledgements} 
We would like to express sincere thanks to the anonymous referee with whom we had a comprehensive and illuminating discourse which allowed us to improve the clarity and veracity of our paper to the current state. TD acknowledges funding by the Deutsches Zentrum f\"ur Luft- und Raumfahrt contract 50\,QR\,1903. AI acknowledges support from the Royal Society. AJ acknowledges partial funding from the European Space Agency (ESA) under partnership agreement 4000133194/20/NL/MH/hm between ESA and FAU Erlangen-N\"urnberg. JAG acknowledges support from an Alexander von Humboldt fellowship. We thank John E.~Davis for the development of the \textsc{SLxfig} module used to prepare the figures in this paper. This research has made use of ISIS functions provided by ECAP/Remeis observatory and MIT (http://www.sternwarte.uni-erlangen.de/isis/).

\section*{Data Availability}
The data underlying this article will be shared on reasonable request to the corresponding author.

\bibliographystyle{mn2e_williams} 
\bibliography{mnemonic,mn_abbrv,local,additional}

\appendix

\section{Derivation of the Returning Photon Flux $\Fret(\Einc, \rinc)$}

In this appendix we present a detailed derivation of the observed returning photon flux $\Fret(\rinc, \Einc)$.

\subsection{Basic Ray-Tracing Equations}
\label{sec:appendix-ray-tracing}

For our calculations we use the line element of the Kerr metric in Boyer-Lindquist coordinates, which is given by \citep{Kerr1963,Bardeen1972}
\begin{equation}
    ds^2 = e^{-2\nu} dt^2 + e^{2\psi}(d\phi - \omega dt)^2
    + \Delta^{-1}\Sigma dr^2  + \Sigma d\theta^2
\end{equation}
where
\begin{align}
    e^{2\nu} &= A^{-1}\Delta\Sigma \\
    e^{2\psi} &= \sin^2\theta A \Sigma^{-1} \\
    \omega &= 2arA^{-1} \\
    A &= (r^2 + a^2)^2 - a^2 \Delta \sin^2\theta \\
    \Delta &= r^2 + a^2 - 2r \\
    \Sigma &= r^2 + a^2\cos^2\theta
\end{align}
Note that all equations in this paper are given in units of $G \equiv M \equiv c \equiv 1$. Assuming that the accretion disk is geometrically thin, it is made of particles orbiting the black hole on circular trajectories in the equatorial plane. It can be shown \citep[see][]{Bardeen1972,Cunningham1975} that the accretion disk then has the 4-velocity 
\begin{align} \label{eq:4veloc}
    u^\mu = u^t (\partial_t + \Omega \partial_\phi) 
\end{align}
with
\begin{align}
    u^t &= e^{-\nu} \left[1-(V^{(\phi)})^2\right]^{-1/2} = e^{-\nu} \Gamma^{(\phi)}  \quad.
\end{align}
Here, $V^{(\phi)}$ is the 3-velocity with respect to the locally non-rotating frame (LNRF) in the direction of the movement of the accretion disk and is given by 
\begin{equation}
 V^{(\phi)} = (\Omega - \omega) e^{\psi-\nu}
\end{equation}
and therefore 
\begin{equation}\label{eq:lorentz}
  \Gamma^{(\phi)}=[1-(V^{(\phi)})^2]^{-1/2}
\end{equation}
is the Lorentz factor for this movement. The coordinate angular velocity, $\Omega$, is given by
\begin{equation}
  \Omega = \frac{1}{r^{3/2} + |a|} \quad.
\end{equation}
The momentum of a null geodesic $p_\mu$, i.e., of a photon, is given by
\begin{align}
    p_t &= -E \\
    p_\phi &= E \lambda \\
    p_r &= \pm E \sqrt{V_r} \Delta^{-1} \\
    p_\theta &= \pm E \sqrt{V_\theta} 
\end{align}
where
\begin{align}
    V_r &= (r^2 + a^2 - a\lambda)^2 - \Delta[(\lambda-a)^2 + q^2]  \\
    V_\theta &= q^2 - \cos^2\theta\left[\frac{\lambda^2}{\sin^2\theta} - a^2 \right] \quad,
\end{align}
where $E$, $\lambda$, and $q$, are the well known constants of motion \citep{Carter1968,Bardeen1972,Cunningham1975}.

\subsection{Locally non-rotating frame and fluid frame}

In order to calculate the incident returning flux (Appendix~\ref{sec:appendix-tf}), we need to define two frames, the fluid frame,  $\mathcal{R}$, which is co-rotating with the accretion disk, and the LNRF, $\hat{\mathcal{R}}$ \citep{Bardeen1972}, of a stationary observer. Following \citet{petrucci1997}, the area $d\hat{A}$ of the annulus in the LNRF can be calculated directly from the metric using the coefficients for the LNRF,
\begin{equation} \label{eq:dA_LNRF}
    d\hat{A} = 2 \pi \sqrt{\frac{A}{\Delta}}  \;dr  
    = 2 \pi e^{-\nu} r \;dr \quad.
\end{equation}
The calculation makes use of the fact that the area is located  in the equatorial plane, where $\theta = \pi/2$.

In order to convert the area $d\hat{A}$ from the stationary to the co-rotating $\mathcal{R}$ frame, we can use the covariance of the space-time quadrivolume for inertial frames \citep[see, e.g.,][]{petrucci1997,niedzwiecki2008,kulkarni2011}, such that
\begin{equation}
    dA\,dt = d\hat{A}\,d\hat{t} \quad,
\end{equation}
again using that the computation is done in the equatorial plane.

We can employ the invariance of $dAdt$ to calculate the returning flux, as for deriving the flux we need to calculate the ratio of the product $dAdt$ between the emitter and observer frame (see Appendix~\ref{sec:appendix-tf}). Due to the invariance of $dAdt$ for inertial frames, we can directly evaluate it in the stationary $\mathcal{\hat{R}}$ frame.  Using the area in the LNRF from Eq.~\ref{eq:dA_LNRF} and that $\hat{u}^t = e^{-\nu}$, this ratio is given as
\begin{equation}\label{eq:dAdt}
    \frac{dA\e dt\e}{dA\o dt\o } = \frac{d\hat{A}\e d\hat{t}\e}{d\hat{A}\o d\hat{t}\o } 
     = \frac{\hat{u}^t\e r\e dr\e d\hat{t}\e}{\hat{u}^t\o  r\o dr\o d\hat{t}\o} = 
    \frac{r\e dr\e}{r\o dr\o} \quad,
\end{equation}
where we used that the energy shift between the two stationary frames is given by
\begin{equation}
    \hat{g} = 
    \frac{p_\mu \hat{u}\o^\mu}{p_\mu \hat{u}\e^\mu}
    = \frac{\hat{u}\o^t}{\hat{u}\e^t}
    = \frac{d\hat{t}\e}{d\hat{t}\o} \quad.
\end{equation}

\subsection{Returning Photon Flux}
\label{sec:appendix-tf}

In order to calculate the observed returning photon flux $\Fret(\rinc, \Einc)$, we need to define the specific photon intensity for a flat source in the fluid frame $\mathcal{R}$. The specific photon intensity is defined in the usual way, as number of photons emitted by the source per unit photon energy per unit time per unit cross-sectional area of the source per unit solid angle. We assume that the returning photons are emitted at a disk patch with area $\dif{A_\mathrm{e}}$. Then, following \citet{chandrasekhar1960}, the specific photon intensity emitted by this source is given as
\begin{equation}
    I\e(E\e,\theta\e,\phi\e) = \frac{1}{\mu\e dA\e} \frac{\dif N\e(E\e)}{ \dif{t\e} \dif{E\e} \dif\Omega\e } \quad.
\end{equation}
Here, the factor $1/\mu\e$ is due to the flat surface of the area, as $\mu\e dA\e$ is the area perpendicular to the direction of $d\Omega\e$. Furthermore, $dN\e(E\e)$ is the number of photons with energy in the range $[E\e,E\e+dE\e]$ that are radiated by the source into a bundle with polar angle $[\theta\e,\theta\e+d\theta\e]$ and azimuthal angle $[\phi\e,\phi\e+d\phi\e]$ during a time interval $dt\e$, and $\mu\e=\cos\theta\e$. All quantities are measured in the source frame. For isotropic radiation, we have $I\e(E\e,\theta\e,\phi\e) = I\e(E\e)$. The source is radiating all of its photons over the northern hemisphere ($0 \leq \theta\e \leq \pi/2$). The number of photons radiated over the entire hemisphere per unit source area, per unit time, per unit photon energy is
\begin{equation}
    F\e(E\e) = \int_{\Omega\e} I\e(E\e,\theta\e,\phi\e) \mu\e \dif\Omega\e.
\end{equation}
This is the specific photon flux radiated by the source. For isotropic radiation, this becomes
\begin{equation}
    F\e(E\e) = I\e(E\e) \int_0^{2\pi} \int_0^1 \mu\e ~\dif\mu\e~\dif\phi\e = I\e(E\e) ~\pi.
\end{equation}
Therefore, for isotropic radiation from a flat surface, $I\e(E\e) = F(E\e) / \pi$.

We now assume that a geodesic with initial polar angle $\theta\e$ will hit the centre of a disk patch with surface area $\dif A\o$  at $r\o$. At this point the angle between the geodesic and the normal of the receiver surface is $\theta\o$. The bundle of geodesics that hit the receiver therefore has an initial polar angle in the range $\theta\e$ to $\theta\e+\dif\theta\e$, and the perpendicular surface area of the beam at the point where it hits the receiver is $\dif A\o^\perp = \mu\o \dif A\o$, where $\mu\o=\cos\theta\o$. The number of photons sent down this bundle with initial energy in the range $E\e$ to $E\e+\dif E\e$ during time interval $\dif t\e$ is
\begin{align}
    \dif N_{\rm bundle}(E\e) &= I\e(E\e) \dif \Omega\e \mu\e \dif A\e \dif t\e \dif E\e \\
    &= \frac{F\e(E\e)}{\pi} ~\dif\Omega\e ~\mu\e ~\dif A\e ~\dif t\e ~\dif E\e \quad.
\end{align}
We wish to measure the specific photon flux that is crossing the disk patch $\dif A\o$ in the rest frame, $\dif F\o(E\o)$. This is the number of photons crossing the receiver per unit time per unit energy per unit area, all measured in the observer frame. Photons emitted from the source with energy $E\e$ will have energy $E\o$ by the time they hit the receiver, where $g=E\o/E\e$ (see Eq.~\ref{eq:ener-shift}). Since the time interval $\dif t\e$ in the emitter frame corresponds to time interval $\dif t\o$ and in the observer rest frame, we can write
\begin{align} \label{eq:dFo}
  \dif F\o(E\o) &= \frac{\dif N_{\rm bundle}(E\e)}{\dif A\o ~\dif t\o ~\dif E\o} \\
  &= F\e(E\e)\frac{\dif \Omega\e ~\mu\e}{\pi} \cdot\frac{ ~\dif A\e ~\dif t\e
    ~\dif E\e}{\dif A\o ~\dif t\o ~\dif E\o} \\
  &= F\e(E\o/g)~\frac{ ~\mu\e}{\pi g} \frac{r\e~\dif r\e}{r\o~\dif r\o}~\dif \Omega\e \quad,
\end{align}
where we have used that $\dif E\o / \dif E\e = g$ and Eq.~\ref{eq:dAdt}.

In order to calculate the total returning photon flux incident on the disk patch $\dif A\o$, we need to integrate the above equation over the full emission of the disk.  The integration is performed over radius and the energy shift \gstar, the latter is parametrizing the azimuthal coordinate $\phi$. Following \citet{Cunningham1975}, we can write the solid angle $\dif \Omega\e$ under which a bundle of photons that is emitted at $r\e$ will hit the disk at $r\o$ within $\dif r\o$ and $g^*$ within $\dif g^*$ as
\begin{equation}
   \dif \Omega\e = \left|\frac{\partial \Omega\e(r\o, g^*)}{ \partial(r\o, g^*)} \right|
   \dif r\o \dif g^*
   \quad.
\end{equation}
Integrating Eq.~\ref{eq:dFo} over the disk then results in 
\begin{align}
 F\o(E\o) 
    &= \int_{r_\mathrm{in}}^{r_\mathrm{out}} \int_0^1
        F\e(E\o/g)~\frac{\mu\e}{\pi g} ~ \frac{r\e}{r\o}
        \left|\frac{\partial \Omega\e(r\o, g^*)}{ \partial(r\o,  g^*)} \right|
        \;\dif r\e \dif g^* \\
    &= \int_{r_\mathrm{in}}^{r_\mathrm{out}} \int_0^1 
            F\e(E\o/g,r\e)~\frac{\Tf(r\e, r\o, g^*)}{r\e} \;\dif r\e \dif g^* \;,
\end{align}
where we defined the \emph{flux transfer function},
\begin{equation}
 \Tf(r\e, r\o, g^*) = \frac{\mu\e }{\pi g} \frac{r\e^2}{r\o}~
        \left|\frac{\partial \Omega\e(r\o, g^*)}{ \partial(r\o, g^*)} \right| \;.
\end{equation}

\subsection{Remarks on the invariance of $dAdt$ in the lamp post geometry and previous publications} \label{sec:append-invariance}

Following the early works of \citet{Cunningham1973}, many different ray-tracing calculations were performed with a large variety of applications. In a popular approach to calculate the irradiation of the accretion disk by a lamp post source on the rotational axis, first used by \citet{Wilkins2012a}, but also used in numerous further publications \citep{Dauser2013a,ingram2019}, the invariance of $dAdt$ was not explicitly used. While in principle the usage of the invariance is not necessary, it has to be invariant nonetheless. In the following, we will show that the approach chosen in these publications is not generally valid, but only in the case of the lamp post geometry. This fact, however, is not explicitly mentioned in these publications. 

In this approach, the irradiating flux is obtained by calculating isotropically distributed null geodesics and counting them in discrete radial bins $dr$ on the accretion disk. As in our case, the resulting flux is then proportional to $\frac{dN}{dAdtdE}$. In order to convert from the emitter to the observer frame on the disk, these publications use that the energy and time transforms with the energy shift along the geodesic by
\begin{equation}\label{eq:time-gshift}
  g = \frac{E\o}{E\e} = \frac{dt\e}{dt\o} = \frac{p_\mu u^\mu\o}{p_\mu u^\mu\e} \quad,
\end{equation}
 since time transforms along the geodesic as the inverse of the energy shift \citep{schneider1992}. 

However, then the area is not calculated under the condition that the quadrivolume (i.e., also $dAdt$ in this case) is invariant between the stationary and co-rotating frame. These publications calculate the disk patch $dA^*$ in the co-rotating frame by simply using the area in the LNRF (Eq.~\ref{eq:dA_LNRF}) and then boost it into the fluid frame by multiplication with the Lorentz factor $\Gamma^{(\phi)}$ (Eq.~\ref{eq:lorentz}), leading to
\begin{equation}
 dA^* = 2\pi u^t r~\dif r \quad. 
\end{equation}
In order to test the invariance of $dAdt$ in this case, we need to calculate the time transformation from the stationary to the co-rotating frame. As in the calculation of the flux, the time transformation $g = dt\e/dt\o$ (Eq.~\ref{eq:time-gshift}) defined by the null geodesic with $p_\mu$ is used, we need to relate the intervals in LNRF $d\hat{t}\e$ and $d\hat{t}\o$ to this definition. Therefore the transformation of the time interval as seen in the co-rotating $\mathcal{R}$ frame with respect to the time interval of the same geodesic as measured in the LNRF $\mathcal{\hat{R}}$ needs to be calculated using the same null geodesic $p_\mu$, but projected on the 4-velocity of each corresponding frame $\mathcal{R}$ and $\mathcal{\hat{R}}$. This results in the time transformation   
\begin{equation} \label{eq:time-trans}
    \frac{dt}{d\hat{t}} = 
     \frac{p_\mu \hat{u}^\mu}{p_\mu u^\mu }
     = \frac{e^{-\nu}}{u^t(1 - \lambda\Omega)}
     = \frac{1}{\Gamma(1 - \lambda\Omega)} \quad,
\end{equation}
with the 4-velocity with respect to the LNRF of ${u^\mu}=u^t(\partial_t + \Omega \partial_\phi)$ and the one of the stationary observer in the LNRF as $\hat{u}^\mu= \hat{u}\partial_t = 1/\sqrt{-g_{tt}} \partial_t = e^{-\nu} \partial_t$. Checking for the invariance of $dA^*dt$, we immediately see that it is not valid in general, as
\begin{align}\label{eq:non-invariance}
 ¸d\hat{A}d\hat{t} &=  2\pi e^{-\nu} r  ~dr~\frac{d\hat{t}}{dt}dt  \\
 &= 2\pi u^t r (1- \lambda \Omega) ~drdt \\
 &= dA^* ~ (1-\lambda\Omega) ~dt \quad.
\end{align}
It has an additional factor 
\begin{equation}
    1 - \lambda\Omega \quad,
\end{equation}
which only reduces to unity for $\lambda=0$. Note that this additional factor $(1-\lambda\Omega)$ can also be attributed to the time interval, such that Eq.~\ref{eq:non-invariance} yields $d\hat{A} d\hat{t} = dA^* dt^*$, meaning that in this case the time does not transform as the inverse of the energy shift due to this additional factor.

However, as all null geodesics in the lamp post geometry have $\lambda=0$ \citep[see][]{Dauser2013a}, the invariance in Eq.~\ref{eq:non-invariance} is valid also in the above mentioned approach. Therefore the results concerning the lamp post geometry in \citet{Wilkins2012a} and the complete results of \citet{Dauser2013a} and \citet{ingram2019}, which only discuss the lamp post geometry, are correct. However, in none of those publications is this strong requirement of the lamp post geometry explicitly stated. Moreover, in \citet{Wilkins2012a} the disk irradiation by extended primary sources and in \citet{wilkins2020} by returning radiation is calculated in the same way, where generally $\lambda \neq 0$, which therefore leads to wrong results. We note that in this case the results could still be correct in case the appropriate time transformation following $dt\o/dt\e = u^t\e/u^t\o$ is used \citep[see, e.g., ][]{schnittman2013}. From everything stated in those papers, however, it seems that instead the time is transformed by the inverse energy shift (Eq.~\ref{eq:time-gshift}).

\section{Flux Correction Factor} \label{sec:flux-corr-fact}

\begin{figure}
  \centering \includegraphics[width=\columnwidth]{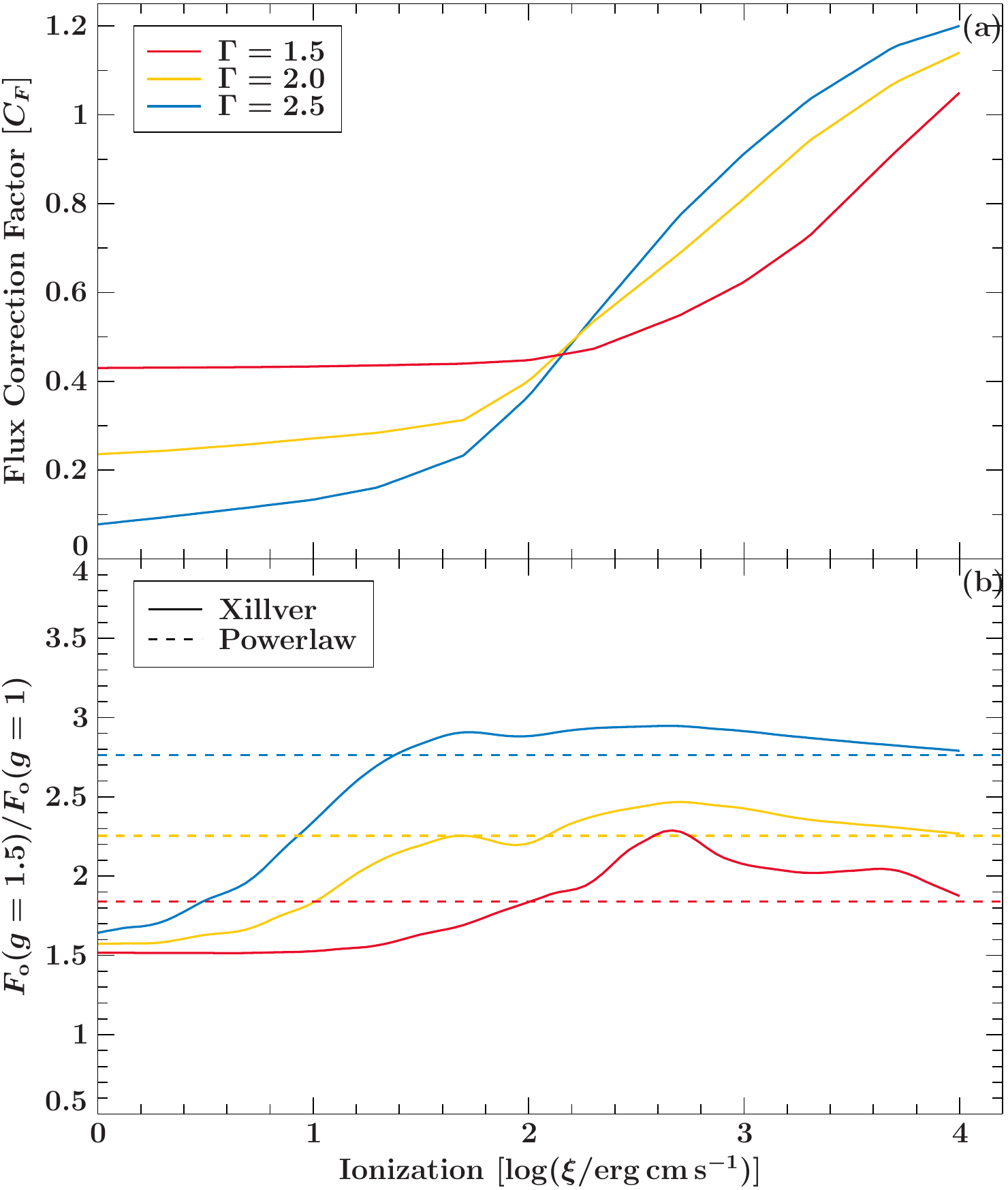}
  \caption{(a) Flux correction factor $C_F$ for standard parameters of the reflection, as it depends on the ionization $\log \xi$ and on the photon index $\Gamma$. (b) shows the flux boost/reduction of the radiation exemplary for an energy shift $g=1.5$ as a function of the ionization, using a \xillver spectrum. Dashed line is the comparison to a power-law spectrum, which transforms as $g^\Gamma$ and therefore does not depend on the ionization. The color indicates the value of $\Gamma$ that has been used.  }
  \label{fig:C_F}
\end{figure}

The flux correction factor $C_F$ (see Sect.~\ref{sec:new-relat-refl-model}) takes into account the fraction of the locally irradiating flux that is is re-processed and emitted in the X-ray energy band. For this purpose we use the range of 0.1--1000\,keV on which the \xillver model is currently defined. The correction factor is defined as the ratio of the locally emitted energy flux with respect to the incident energy flux. We calculate it using \xillver reflection spectra integrated over the emission angles assuming isotropy. The dependency of the flux correction factor on the ionization and the photon index of the irradiating spectrum is shown in Fig.~\ref{fig:C_F}a. 

Figure~\ref{fig:C_F}b shows how the flux boost depends on the shape of the spectrum and the ionization. While for a power law spectrum this flux boost does not depend on the ionization, using a detailed \xillver reflection spectrum the ionization strongly influences how the flux of the spectrum is affected by the energy shift $g$. The flux ratio is approximately calculated by shifting one spectrum by $g=1.5$, the average of the expected value of $g$ (Fig.~\ref{fig:gshift_line}). Then the energy flux is determined for both, the shifted and non-shifted spectrum, in the 0.15--500\,keV band. The reason for the slightly smaller energy range than that used by \xillver\ is to have a common energy band where both spectra are defined, after one is shifted by the factor $g$. In order to allow for model fitting in real-time including returning radiation, the \relxill\ model uses this factor of $F_\mathrm{o}(g=1.5)/F_\mathrm{o}(g=1)$ and interpolates linearly from this for a given energy shift $g$. 

For low ionization the flux boost for a \xillver spectrum is very similar for different values of $\Gamma$. A higher degree of ionization leads to an increase in flux boost, over-shooting the power-law boost, and then again converging towards it. Generally, for $\log(\xi/\mathrm{erg\,cm\,s}^{-1})>2.5$ the difference in flux boost between \xillver and a power law spectrum is small. This can be understood, as for a larger ionization the  reflection spectrum resembles more the irradiating spectrum \citep[see, e.g.,][]{Garcia2013a}. Note that combining the flux correction and the difference in flux boost is close to unity for an ionization of $\log(\xi/\mathrm{erg\,cm\,s}^{-1}) \approx 3$. Therefore the reflection indeed seems to behave as for a "perfect reflector" in terms of its flux.

\section{Comparison with previous ray-tracing calculations} \label{sec:comparison-with-ak00-and-wilkins}

In the following a comparison to two similar ray tracing simulations by \citet{Agol2000a} and \citet{wilkins2020} is presented. While our assumptions are identical to the ones by \citet{wilkins2020}, the calculations from \citet{Agol2000a} we use for comparison were performed for $a=0.9999$. However, comparisons showed that for the same radius, no larger differences are expected to the results with $a=0.998$.

\begin{figure}
\includegraphics[width=\columnwidth]{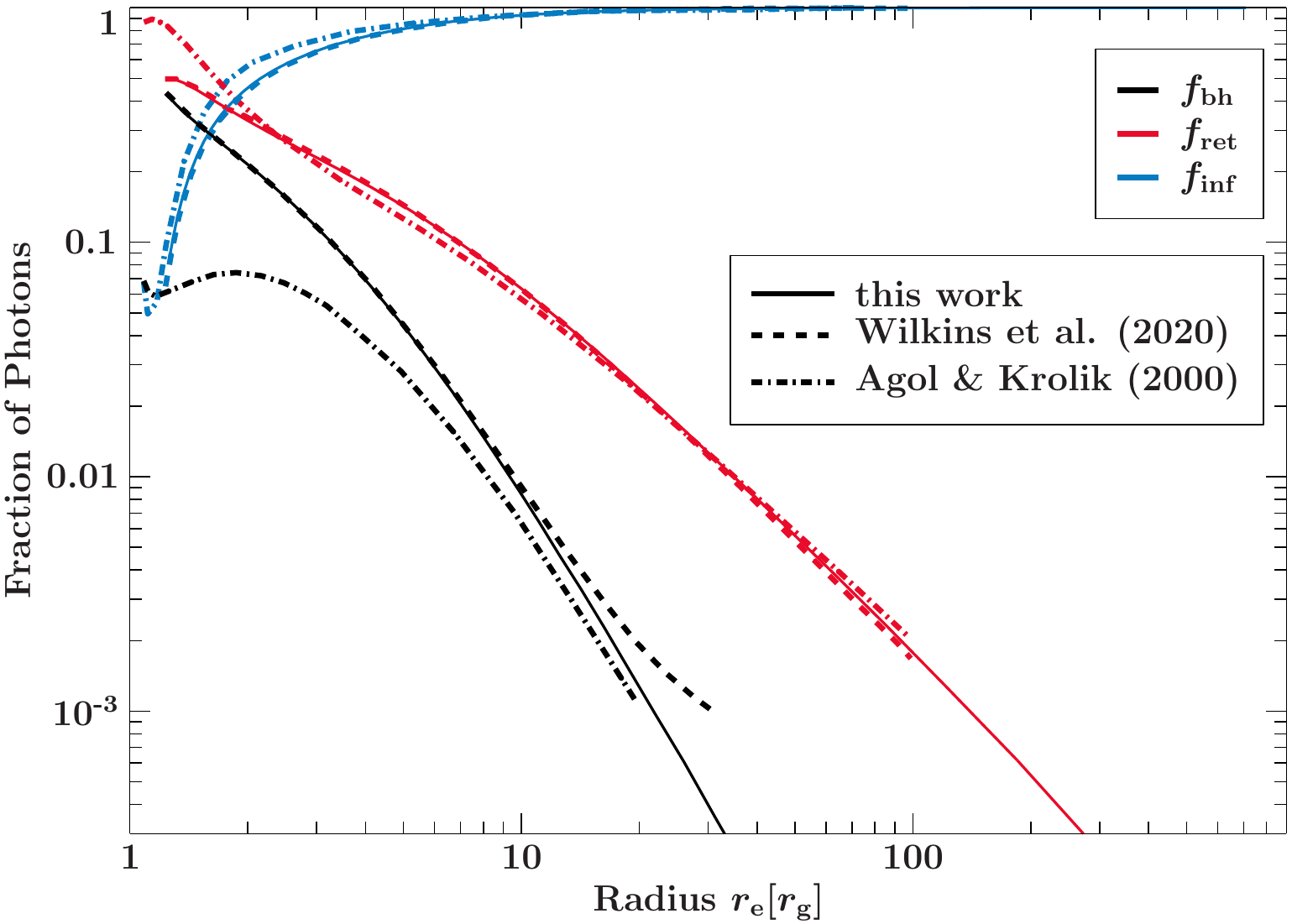}
  \caption{Comparison of our ray-tracing results shown in Fig.~\ref{fig:photon_fate}  with \citet[dashed-dotted]{Agol2000a} and \citet[dashed]{wilkins2020}. }
  \label{fig:comparison_fractions}
\end{figure}

In Fig.~\ref{fig:comparison_fractions} we present the comparison of these results. It can be seen that for large radii there is very good agreement and similar asymptotic behavior between all curves. The differences for larger radii can be understood, as in the work by \citet{wilkins2020} rays are counted as returning only if they are not higher than 500\rg above the disk and counted as escaping for $r>1000\rg$. In the case of \citet{Agol2000a}, there is no mention of where the outer disk ends, but it was likely set at a very large radius. However, for smaller radii there are significant differences. At these radii, our results are very close to the ones produced by \citet{wilkins2020}.

\label{lastpage}

\end{document}